\documentclass[review]{elsarticle}

\usepackage{lineno,hyperref}
\usepackage[nolist]{acronym}
\usepackage{amsmath,xparse,color,mathtools}
\modulolinenumbers[5]
\graphicspath{{artwork/}}

\newcommand{\vb}[1]{\mathbf{#1}}

\NewDocumentCommand{\evalat}{sO{\big}mm}{   
  \IfBooleanTF{#1}                          
   {\mleft. #3 \mright|_{#4}}
   {#3#2|_{#4}}%
}

\begin{acronym}
  \acro{1DOF}{1-degree of freedom}
  \acro{NDOF}{N-degrees of freedom}
  \acro{AVMI}{Added Virtual Mass Incremental}
  \acro{NAVMI}{Nondimensional Added Virtual Mass Incremental}
  \acro{CFD}{Computational Fluid Dynamics}
  \acro{FEA}{Finite Element Analysis}
  \acro{FSI}{Fluid-Structure Interactions}
  \acro{MMS}{Memory Management System}
  \acro{RSI}{Rotor-Stator Interaction}
  \acro{RK4}{Runge-Kutta method}
  \acro{URANS}{Unsteady Reynolds-Averaged Navier–Stokes}
\end{acronym}

\journal{arXiv.org}

\biboptions{authoryear}

\begin{document}

\begin{frontmatter}

\title{Modal analysis of a spinning disk in a dense fluid as a model for high head hydraulic turbines}

\author[1]{Max Louyot}
\ead{max.louyot@gmail.com}
\author[2]{Bernd Nennemann}
\ead{bernd.nennemann@andritz.com}
\author[2]{Christine Monette}
\ead{christine.monette@andritz.com}
\author[1]{Frederick P. Gosselin}
\ead{frederick.gosselin@polymtl.ca}

\address[1]{D{\'e}partement de G{\'e}nie M{\'e}canique, Laboratory for Multiscale Mechanics (LM2), Polytechnique Montr{\'e}al, Montreal, QC, Canada}
\address[2]{Andritz Hydro Canada Inc., Pointe Claire, QC, Canada}

\begin{abstract}
In high head Francis turbines and pump-turbines in particular, Rotor Stator Interaction (RSI) is an unavoidable source of excitation that needs to be predicted accurately. Precise knowledge of turbine dynamic characteristics, notably the variation of the rotor natural frequencies with rotation speed and added mass of the surrounding water, is essential to assess potential resonance and resulting amplification of vibrations. In these machines, the disk-like structures of the runner crown and band as well as the head cover and bottom ring give rise to the emergence of diametrical modes and a mode split phenomenon for which no efficient prediction method exists to date. Fully coupled Fluid-Structure Interaction (FSI) methods are too computationally expensive; hence, we seek a simplified modelling tool for the design and the expected-life prediction of these turbines.\\
We present the development of both an analytical modal analysis based on the assumed mode approach and potential flow theory, and a modal force Computational Fluid Dynamics (CFD) approach for rotating disks in dense fluid. Both methods accurately predict the natural frequency split as well as the natural frequency drift within $7.9\%$ of the values measured experimentally. The analytical model explains how mode split and drift are respectively caused by linear and quadratic dependence of the added mass with relative circumferential velocity between flexural waves and fluid rotation.
\end{abstract}


\end{frontmatter}

\linenumbers

\section{Introduction}

The design of modern hydroelectric turbines aims at near perfect efficiency while minimizing production costs. In this context, precise knowledge of the turbine dynamical characteristics, notably the variation of the rotor natural frequencies with rotation speed and added mass of the surrounding water, is essential to assess potential resonance and resulting amplification of vibrations. Turbine design requires fast and numerically efficient frequency identification methods taking into account these effects.

The manifold complexity of hydraulic turbines makes them subject to numerous physical effects. Resonance of structures coupled to excitation sources can lead to severe fatigue damage, that can result in the loss of hydraulic runner blades if it is not avoided \citep{coutu2004,coutu2008,liu2016}. Typically, the extra stresses due to daily start-stop cycles that turbines now undergo can initiate cracks in the blades \citep{huth2005,trivedi2017}. The intricacy of hydraulic turbine design and the numerous possible vibration sources in these machines make the runner reliability a challenging and critical design criteria \citep{dorfler2012,presas2019}. In particular in high head Francis turbines and pump-turbines, \ac{RSI} are an unavoidable source of excitation that needs to be predicted accurately \citep{ohashi1994,dorfler2012,walton2016} in order to ensure the designed geometry is suitable before its manufacturing.

Fully understanding the physics at stake is crucial in order to develop design analysis methods that allow for a design process leading to safe and robust machines in an economical period of time. Critical rotation speeds of spinning structures trigger unstable regimes, such as flutter \citep{adams1987,renshaw1998,kim2000}. Each structural mode can typically be stabilized by increasing the stiffness. The fluid filled spaces between rotor and casing also modify this threshold \citep{huang1995,amabili1996_3}, especially with narrow gaps, like in hydraulic turbines. Additionally, the coupling between the acoustical and structural natural frequencies, the radial gap, the geometrical asymmetries, and the fluid rotation are all parameters upon which rotating structure stability relies \citep{kang2006_1,kang2006_2}. The natural frequencies of runners are affected by equally numerous parameters including their rotational velocity and the influence of surrounding water \citep{egusquiza2016}. Acoustical natural frequencies can decrease natural frequencies by up to 25\% if coupling occurs \citep{bossio2017}. Such frequencies are usually higher than rotation speeds for high head turbines. This review demonstrates the complexity of both runner physics and geometry. Hence, the untangling of physical phenomenon and relevant parameters of influence on runner natural frequencies requires a simplified approach.

\begin{figure}[t]
    \centering
    \includegraphics[width=\textwidth]{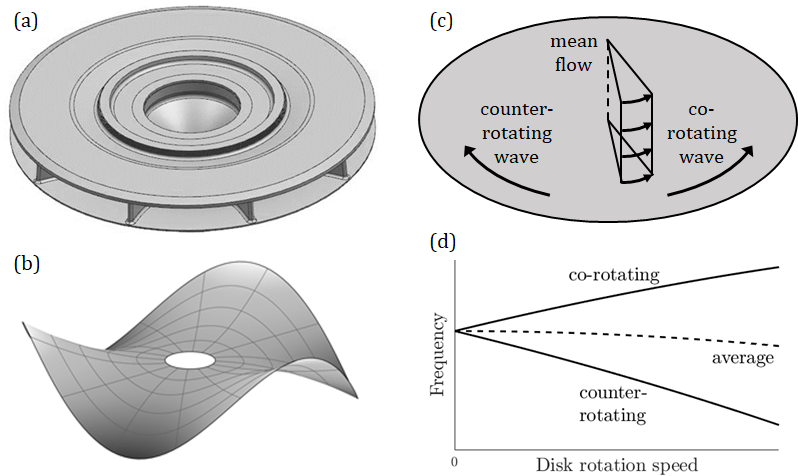}
    \caption{(a) A pump-turbine runner geometry. (b) High head Francis and pump-turbine runners have disk-like modeshapes, characterized by their numbers $n$ and $s$ of nodal diameters and nodal circles respectively ($n=3$, $s=0$ is presented). (c) Each disk mode is composed of a co-rotating and a counter-rotating wave, with respect to the fluid rotation relative to the disk. (d) Co- and counter-rotating wave frequencies evolve with the disk rotation speed in dense fluid: the split between the two increases, while the average value decreases.}
    \label{fig:TurbDiskSplit}
\end{figure}

If efficient frequency prediction methods exist for the dynamical response of low to medium head Francis runners \citep{coutu2008} and of Kaplan turbines \citep{soltani2017,soltani2019}, this is not the case for high head turbines and in particular pump-turbine runners. Vibration modeshapes of these runners are different due to the disk-like structures of the runner crown and band, which give rise to diametrical modes, as shown in \mbox{Figure \ref{fig:TurbDiskSplit}}(a-b). Modeshapes of eigenfrequencies typically below 450 Hz are disk-like modes \citep{egusquiza2016}, validating the disk representation of high head turbine runners for a large range of rotation speeds. Hence, the present work studies an idealized rotating plate in dense fluid. Beyond simplifying the rotor geometry to that of a disk, we can further idealize the problem with the following assumptions:
\begin{enumerate}
    \item The rotation speed range considered for hydraulic turbine applications is low enough to neglect centrifugal forces in the disk.
    \item For the selected potential flow approach, the fluid surrounding the disk is considered inviscid and adiabatic.
    \item From the rotating disk reference frame, based on work from \citet{poncet2005}, the fluid is entrained to a solid-body motion at a mean velocity equal to a fraction of that of the disk.
    \item The disk modes in water are the same as those of the disk in vacuum. This was confirmed by \citet{kwak1991} with the Rayleigh-Ritz method.
    \item We consider small amplitude deformations of the disk in order to remain in the frame of linear perturbation analysis.
\end{enumerate}
These assumptions allow us to orient our literature review.

\citet{leissa1969} compiles analytical solutions for annular plate modes and various boundary condition sets, using previous work from \citet{southwell1922} and \citet{vogel1965} including geometries and conditions of particular interest for the present work.

Submerged structure resonance frequencies are shifted by the surrounding water effect \citep{liang2007,ostby2019}. Part of the fluid vibrates with the structure, adding mass to the system. The plate rotation in water additionally triggers a particular resonance mechanism, first described by \citet{kubota1991}: unlike in air, forward and backward travelling waves on the disk surface trigger the so-called mode split, as a mode can be excited with two different frequencies, as shown in Figure \ref{fig:TurbDiskSplit}(c-d). However, there is little information on the physical phenomenon itself. \citet{presas2014,presas2015,presas2016} along with \citet{valentin2014} analytically and experimentally studied the natural frequencies of a submerged and confined rotating disk, as well as the influence of the rotation speed and of the axial gap length. Their experimental setup consists of a rotating disk excited with a piezoelectric patch, surrounded by air or water in a fully-rigid casing. The disk response was measured with accelerometers mounted on its surface, allowing the detection of the first structural modeshapes and associated frequencies. They showed that reducing the axial gap increases the mode split effect and decreases natural frequencies.

Various methods exist to study the vibrations of rotating structures. While considering coupled structural eigenmodes may prove close to reality, it can also be difficult to implement. Modal analysis only deals with independent modes, greatly simplifying the problem to solve. \citet{ahn1998} analytically studied the steady-state modal response of an excited rotating disk, and identified the modes linked to forward and backward travelling waves, along with their associated frequencies. \citet{renshaw1994} identified the ratio of the fluid and plate densities to be one of the main influencing parameters: mode split arises when the structure interacts with dense fluids such as water. \citet{kwak1991} and \citet{amabili1996_2} analytically analyzed stationary circular plates coupled with water. They used the assumed mode approximation and worked with a potential flow. The assumed shape of the potential function must satisfy the boundary conditions; it provides spatial and temporal information on the velocity field. Then, assessing the \ac{NAVMI} factor from the potential flow and imposed modeshape, they linked the natural frequency in vacuum to that in dense fluid. This factor represents the ratio between reference kinetic energies of the disk and fluid. \citet{amabili1996_1} performed the same analysis on standing annular plates, proving its applicability to the disk considered in our work. The model developed by \citet{presas2016} includes the relative rotation between the fluid and disk, which gives rise to mode split. It expresses the axial deformation at a characteristic radius $r_0$, and assumes that natural frequencies can be determined only from the disk dynamical information at $r=r_0$. The axial gap dimensions are taken into account through the boundary conditions. Unfortunately, all of these models still lack an exhaustive understanding of the physical nature of mode split and present a non-negligible relative error on the frequency predictions.

Analytical solutions cannot be established for hydraulic turbine complex geometries. Numerical \ac{FSI} models using \ac{FEA} are powerful tools for the structural design analysis of these machines \citep{dompierre2010,hubner2016}. This method is applicable to natural frequency prediction of standing circular plates \citep{hengstler2013}, disk-fluid-disk systems, rotor-stator systems \citep{specker2016,weder2016,weder2019}, and rotating disks in fluid \citep{weber2015}. This shows that the application range of numerical \ac{FSI} is extremely broad, and widely used to deal with disk frequency prediction. However, these fully-coupled simulations are computationally expensive \citep{hubner2016,nennemann2016,biner2017} and are not a convenient tool to evaluate runner natural frequencies in flow in the preliminary design stage \citep{weber2015}. They also present stability issues when the fluid-to-structure density ratio increases \citep{wong2013}, typically when the fluid is water. Therefore, a simplified approach to model high head Francis and pump-turbines would be a powerful tool in the scope of our study. Several authors suggested using a faster approach than fully-coupled \ac{FSI}, such as transient flow \citep{soltani2017} or structural-acoustical methods \citep{valentin2016,escaler2018}. But because of both its simplicity and efficiency for submerged and confined rotating disks, the most promising alternative is a modal force approach \citep{nennemann2016,presas2016,biner2017}, for which an arbitrary number of modes considered separately are imposed on the disk surface. A time discretization of structural equations allows the computation of the displacement according to the surrounding fluid parameters and pressure field obtained with \ac{CFD}. In order to numerically capture the mode split on rotating submerged disks, our work builds on the \ac{1DOF} oscillator modal force \ac{CFD} model developed by \citet{monette2014} and \citet{nennemann2016}. This model was originally used to predict added stiffness and damping of runner blades in flowing water.

Here we present the development of both an analytical modal analysis based on the assumed mode approach and potential flow theory, and a modal force \ac{CFD} approach for rotating disks in dense fluid. Both methods accurately predict the natural frequency split and drift that are observed experimentally by \citet{presas2016}. Insight into the physical origin of the mode split is also given. Both models are validated by comparison with available experimental data.
\section{Methodology}\label{sec:method}

In this section we detail the development of the structural and fluid equations leading to the analytical modal analysis, and the modal force \ac{CFD} approaches. The disk material, geometry and rotation speed, as well as the fluid and casing properties are taken into account. The radial gap is only considered in the numerical model.

\begin{figure}[t]
    \centering
    \includegraphics[scale=0.9]{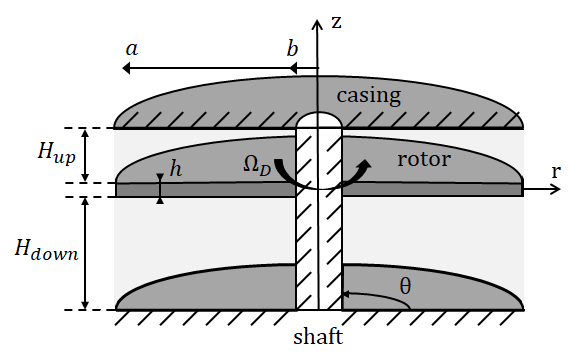}
    \caption{Studied geometry in \S \ref{sec:struct} and \S \ref{sec:ana_method}: the disk with angular speed $\Omega_D$, outer radius $a$, inner radius $b$ and thickness $h$ is confined in a rigid casing of height $H_{up}+H_{down}$, filled with water. The disk is clamped to the shaft on its inner radius and free outside.}
    \label{fig:Geom}
\end{figure}

The disk is a rotating annular plate of density $\rho_D$, outer radius $a$, inner radius $b$, thickness $h$ and angular speed $\Omega_D$ in the stationary reference frame. We use the cylindrical coordinates $(r,\theta,z)$ in the stationary reference frame, and the origin is taken at the center of the disk. The rigid casing of height $H_{up}+H_{down}$ is filled with a liquid of density $\rho_F$. The gap between the disk and the top of the casing is of length $H_{up}$, while the gap between the disk and the bottom of the casing is of length $H_{down}$. \mbox{Figure \ref{fig:Geom}} presents the modeled geometry.

\subsection{Structural model}\label{sec:struct}

Let us first establish the structural equations upon which rely both the analytical and numerical methods. According to linear classical plate theory, the vertical displacement $w$ of an annular plate is given by \citet{leissa1969} as
\begin{equation}\label{eq:wDisk}
    D\nabla^4w+\rho_Dh\frac{\partial^2w}{\partial t^2}=P(r,\theta,t)\,,
\end{equation}
where $t$ is the elapsed time, $P(r,\theta,t)$ is the pressure field applied to the plate, $\nabla^4=\nabla^2\nabla^2$ with $\nabla^2$ the Laplacian operator, and
\begin{equation}
    D=\frac{Eh^3}{12(1-\nu^2)}\,,
\end{equation}
is the disk flexural rigidity, where $E$ is Young's modulus and $\nu$ is Poisson's ratio of the disk material, most likely stainless steel for hydraulic turbine applications. As turbine runners are typically continuous steel structures vibrating in water with negligible material damping and no frictional damping, most of their damping is flow-induced \citep{gauthier2017}, hence structural damping is neglected. We assume Fourier components in $\theta$, where $n$ (respectively $s$) is the number of nodal diameters (respectively nodal circles), and $W_{ns}$ is the associated modeshape in vacuum ($P=0$). The vertical displacement in vacuum is composed with all modeshapes:
\begin{equation}
    w(r,\theta,t)=\sum_{n=0}^{\infty}\sum_{s=0}^{\infty}W_{ns}(r,\theta)e^{i\omega_vt}\,,
\end{equation}
where $\omega_v$ is the natural angular frequency in vacuum. In order to determine the shape of $W_{ns}$, we introduce the parameter $k_{ns}$ defined as
\begin{equation}\label{eq:k}
    k_{ns}^4=\frac{\rho_D h\omega_v^2}{D}\,,
\end{equation}
which yields the information on $\omega_v$. For annular plates, \citet{leissa1969} provides the modeshape for a single mode in the form
\begin{equation}\label{eq:modeshape}
    W_{ns}(r,\theta)=\psi_{ns}(r)\,e^{in\theta}\,,
\end{equation}
where
\begin{equation}\label{eq:modeshape_r}
    \psi_{ns}(r)=A_nJ_n(k_{ns}r)+B_nY_n(k_{ns}r)+C_nI_n(k_{ns}r)+D_nK_n(k_{ns}r)\,,
\end{equation}
where $J_n,Y_n,I_n,K_n$ are the Bessel functions of first and second kinds, and the modified Bessel functions of first and second kinds respectively, and $A_n,B_n,C_n,$ $D_n$ are coefficients determined with the boundary conditions within a multiplying factor. It should be noted that the complex conjugate part is omitted in order to lighten the equations. The boundary conditions for the free-clamped annular plate are also given by \citet{leissa1969}:
\begin{align}\label{eq:BC1}
    \evalat{W_{ns}(r,\theta)}{r=b}&=0\,,&\evalat[\bigg]{\frac{\partial W_{ns}(r,\theta)}{\partial r}}{r=b}&=0\,,\\\label{eq:BC2}
    \evalat{V_r(r,\theta)}{r=a}&=0\,,&\evalat{M_r(r,\theta)}{r=a}&=0\,,
\end{align}
where $V_r$ is the radial Kelvin-Kirchhoff edge reaction and $M_r$ is the bending moment:
\begin{align}
    V_r&=-D\bigg[\frac{\partial}{\partial r}(\nabla^2W_{ns}(r,\theta))+(1-\nu)\frac{1}{r}\frac{\partial^2}{\partial\theta\partial r}\bigg(\frac{1}{r}\frac{\partial W_{ns}(r,\theta)}{\partial\theta}\bigg)\bigg]\,,\\
    M_r&=-D\bigg[\frac{\partial^2W_{ns}(r,\theta)}{\partial r^2}+\nu\bigg(\frac{1}{r}\frac{\partial W_{ns}(r,\theta)}{\partial r}+\frac{1}{r^2}\frac{\partial^2W_{ns}(r,\theta)}{\partial\theta^2}\bigg)\bigg]\,.
\end{align}

Substituting Eqs.\ (\ref{eq:modeshape}-\ref{eq:modeshape_r}) in Eqs.\ (\ref{eq:BC1}-\ref{eq:BC2}) results in an eigenvalue problem of the form:
\begin{equation}
    \vb{M}\cdot\vb{X_n}=\begin{bmatrix}
    a_i & \cdots & d_i\\
    \vdots & \ddots & \vdots\\
    a_{iv} & \cdots & d_{iv}
    \end{bmatrix}\cdot\begin{bmatrix}
    A_n\\B_n\\C_n\\D_n
    \end{bmatrix}=\vb{0}\,,
\end{equation}
where the matrix coefficients $a_i\ ...\ d_{iv}$ are given by the developed boundary conditions. We then have $det(\vb{M})=0$ as a necessary condition for the system to be solved. Solving $det(\vb{M})=0$ has an infinite number of solutions which correspond to the values of $k_n$ for any number of nodal circles $s$. This cannot be solved analytically, and $k_{ns}$ must be evaluated numerically. After eliminating the trivial solution $k_{ns}=0$ and moving upward, the $j^{\rm{th}}$ solution that nullifies $det(\vb{M})$ corresponds to $s=j-1$. Then, in order to close the system of equations, we arbitrarily choose the value of $A_n$ and then solve the system to get the values of $B_n,C_n,D_n$, and hence the associated modeshape. These modeshapes naturally form an orthogonal base, and we additionally choose $A_n$ to make it orthonormal.

Let us apply the Galerkin method to the disk structure. The displacement is approximated with a \mbox{discrete sum:}
\begin{equation}\label{eq:galerkin1}
    w(r,\theta,t)\approx w_N(r,\theta,t)=\sum_j^N\phi_j(r,\theta)q_j(t)\,,
\end{equation}
where $N$ is the number of modes of different combinations of nodal diameters and nodal circles $ns$ considered, $\phi_j$ are test functions that satisfy the boundary conditions and $q_j$ are the generalized coordinates. Here, the test functions are chosen to correspond to the orthonormal modeshapes: \mbox{$\phi_j(r,\theta)=W_j(r,\theta)$} defined in Eq.\ (\ref{eq:modeshape}). Calculation shows that this choice implies $\nabla^4W_j=k_j^4W_j$. Hence, substituting $w$ with $w_N$ \mbox{in Eq.\ (\ref{eq:wDisk})} leads to
\begin{equation}
    D\nabla^4w_N+\rho_Dh\frac{\partial^2w_N}{\partial t^2}-P(r,\theta,t)=\mathcal{R}\,,
\end{equation}
where $\mathcal{R}$ is the residual. Expanding the sum of Eq.\ (\ref{eq:galerkin1}):
\begin{equation}\label{eq:galerkin2}
    \sum_j^N\big[\rho_Dh\ddot{q}_j(t)+Dk^4_jq_j(t)\big]W_j(r,\theta)-P(r,\theta,t)=\mathcal{R}\,,
\end{equation}
where $\ddot{q}_j=\partial^2q_j/\partial t^2$. According to the Galerkin method:
\begin{equation}
    \int_{r=b}^a\int_{\theta=0}^{2\pi}\mathcal{R}W_i(r,\theta)\,rdrd\theta=0\ \forall i\leq N\,.
\end{equation}
Let us recall that because the modes form an orthonormal base,
\begin{equation}\label{eq:norm}
    \int_{r=b}^a\int_{\theta=0}^{2\pi}W_i(r,\theta)W_j(r,\theta)r\,drd\theta=2\pi(a^2-b^2)\delta_{ij}\,,
\end{equation}
where $\delta_{ij}$ is the Kronecker symbol. Therefore, multiplying Eq.\ (\ref{eq:galerkin2}) by $W_i$ and integrating over the annular plate surface, we obtain
\begin{equation}\label{eq:galerkin3}
    2\pi(a^2-b^2)\sum_j^N\big[\rho_D h\ddot{q_j}(t)+Dk_j^4q_j(t)\big]\delta_{ij}=\widetilde{P}_i(t)\,,
\end{equation}
where
\begin{equation}
    \widetilde{P}_i(t)=\int_{r=b}^a\int_{\theta=0}^{2\pi}P(r,\theta,t)W_i(r,\theta)\,rdrd\theta\,.
\end{equation}
Let us recall that this is true for any number of considered modes $N$ chosen arbitrarily. Observing that the equations from the obtained system are decoupled, we can then generalize by replacing $j$ by any combination of nodal diameters and nodal circles $ns$ in Eq.\ (\ref{eq:galerkin3}):
\begin{equation}\label{eq:galerkin4}
    2\pi(a^2-b^2)\big[\rho_Dh\ddot{q_{ns}}(t)+Dk_{ns}^4q_{ns}(t)\big]=\widetilde{P}_{ns}(t)\,.
\end{equation}
Hence, any point on the disk can be assimilated to a \ac{1DOF} mass-spring system with vertical motion for any given mode. The vertical displacement $w$ of this point follows
\begin{equation}\label{eq:1dof}
    M\ddot{w}(t)+K_{ns}w(t)=\widetilde{P}_{ns}(t)\,,
\end{equation}
where $M=2\pi(a^2-b^2)\rho_Dh$ and $K_{ns}=2\pi(a^2-b^2)Dk_{ns}^4$ are the modal structural mass and stiffness respectively. It can be seen that the structural modal mass only depends on the mass and geometry of the disk, regardless of the selected mode; while the structural modal stiffness depends through $k_{ns}$ on the number of nodal diameters and circles of the mode. The disk natural angular frequency in vacuum can be assessed with Eq.\ (\ref{eq:1dof}):
\begin{equation}\label{eq:freqvac}
    \omega_v=\sqrt{\frac{K_{ns}}{M}}=k_{ns}^2\sqrt{\frac{D}{\rho_Dh}}\,.
\end{equation}

\subsection{\ac{FSI} analytical model}\label{sec:ana_method}

In this section we develop an analytical model based on modal analysis, for the prediction of rotating disk natural frequencies. Let us recall that modeshapes are considered identical in air and in water. The axial confinement of the fluid is considered, and imposes the radial boundary conditions.

At this point, because $k_{ns}$ is known, we have an expression for any annular plate modeshape given by Eq.\ (\ref{eq:modeshape}), and associated angular frequency in vacuum given by Eq.\ (\ref{eq:freqvac}). As we only consider a single mode with $n$ nodal diameters, \mbox{Eq.\ (\ref{eq:galerkin1})} translates into
\begin{equation}\label{eq:w_1dof}
    w(r,\theta,t)=W_{ns}(r,\theta)g(t)\,,
\end{equation}
with
\begin{equation}\label{eq:temp_part}
    g(t)=e^{i\omega t}\,,
\end{equation}
and $\omega$ is the actual angular frequency of the structure for the considered mode. This harmonic motion assumption allows us to perform an analytical modal analysis to extract the natural frequencies of the coupled fluid-structure system.

We consider the fluid velocity $\vb{V}$ in the surrounding fluid, the form of which is assumed to be the sum of a mean flow component $\vb{V_0}$, associated with the solid body rotation, and of an oscillatory component $\vb{v}$, associated with the transverse disk motion:
\begin{equation}
    \vb{V}(r,\theta,z,t)=\vb{V_0}(r)+\vb{v}(r,\theta,z,t)\,.
\end{equation}
The solid body rotation is described by
\begin{equation}\label{eq:solidbodymotion}
    \vb{V_0}=\{0,(1-K)r\Omega_D,0\}^\textrm{T}\,,
\end{equation}
in the cylindrical reference frame, where $K$ is the average entrainment coefficient, the value of which will be discussed \mbox{in \S\ref{sec:numres}}. This coefficient verifies $K=\Omega_F/\Omega_D$, where $\Omega_F$ is the fluid angular speed in the stationary reference frame.

We associate the oscillatory component $\vb{v}$ to the flow potential $\Phi$:
\begin{equation}\label{eq:velocity}
    \vb{v}=\nabla\Phi\,,
\end{equation}
where we assume the shape of $\Phi$ to be
\begin{equation}\label{eq:Phi}
    \Phi(r,\theta,z,t)=\phi(r,z)e^{in\theta}\dot{g}(t)\,,
\end{equation}
where $\dot{g}=dg/dt$, $g$ is defined in Eq.\ (\ref{eq:temp_part}) and $\phi$ has to be determined. By convention, $n>0$ (respectively $n<0$) characterizes co-rotating waves (respectively counter-rotating waves) relative to the rotating disk. This flow obeys the Laplace equation, which implies
\begin{equation}\label{eq:Laplace_Phi}
    \nabla^2\Phi=\frac{\partial^2\Phi}{\partial r^2}+\frac{1}{r}\frac{\partial\Phi}{\partial r}+\frac{1}{r^2}\frac{\partial^2\Phi}{\partial\theta^2}+\frac{\partial^2\Phi}{\partial z^2}=0\,.
\end{equation}
Substituting $\Phi$ for its expression in Eq.\ (\ref{eq:Phi}) yields the equation solved by $\phi$:
\begin{equation}\label{eq:Laplace_phi}
    \frac{\partial^2\phi}{\partial r^2}+\frac{1}{r}\frac{\partial\phi}{\partial r}-\frac{n^2}{r^2}\phi+\frac{\partial^2\phi}{\partial z^2}=0\,.
\end{equation}
The non-penetration boundary conditions on the top and bottom casing and disk surfaces imply
\begin{align}\label{eq:BCcasing}
    &\evalat[\bigg]{\frac{\partial\phi}{\partial z}}{z=H_{up}}=0\,,\hspace{1cm}\evalat[\bigg]{\frac{\partial\phi}{\partial z}}{z=H_{down}}=0\,,\\
    &\evalat[\bigg]{\frac{\partial\phi}{\partial z}}{z=0}=\frac{Dw}{Dt}=\frac{\partial w}{\partial t}+\frac{\mathrm{V}_{0\theta}}{r}\frac{\partial w}{\partial\theta}\,,\label{eq:BCz0_1}
\end{align}
where $D/Dt$ is the material derivative, and $\mathrm{V}_{0\theta}$ is the tangential velocity in the fluid reference frame. Then, replacing $w$ in Eq.\ (\ref{eq:BCz0_1}) by its expression given by \mbox{Eqs.\ (\ref{eq:modeshape},\,\ref{eq:w_1dof})} yields
\begin{equation}
    \evalat[\bigg]{\frac{\partial \phi}{\partial z}}{z=0}=\Big(1+\frac{n}{\omega}\frac{\mathrm{V}_{0\theta}}{r}\Big)\psi_{ns}(r)\,.
\end{equation}
Substituting the tangential velocity $\mathrm{V}_{0\theta}$, given by Eq.\ (\ref{eq:solidbodymotion}), finally provides
\begin{equation}\label{eq:BCdisk}
    \evalat[\bigg]{\frac{\partial \phi}{\partial z}}{z=0}=\Big(1+\frac{n\Omega_{D/F}}{\omega}\Big)\psi_{ns}(r)\,,
\end{equation}
where $\Omega_{D/F}=\Omega_D-\Omega_F=(1-K)\Omega_D$ is the disk angular speed with respect to the fluid.

Similarly to \citet{amabili1996_1}, we introduce the Hankel transform and the inverse Hankel transform based on Bessel functions:
\begin{align}\label{eq:Hankel}
    \overline{\phi}(\xi,z)&=\int_0^{\infty}r\phi(r,z)J_n(\xi r)\,dr\,,\\
    \phi(r,z)&=\int_0^{\infty}\xi\overline{\phi}(\xi,z)J_n(\xi r)\,d\xi\,.\label{eq:invHankel}
\end{align}
Multiplying Eq.\ (\ref{eq:Laplace_phi}) by $rJ_n(\xi r)$ and integrating over the radius leads to
\begin{equation}\label{eq:Laplace_int}
    \int_0^{\infty}r\frac{\partial^2\phi}{\partial z^2}J_n(\xi r)\,dr=-\int_0^{\infty}r\Big(\frac{\partial^2\phi}{\partial r^2}+\frac{1}{r}\frac{\partial\phi}{\partial r}-\frac{n^2}{r^2}\phi\Big)J_n(\xi r)\,dr\,.
\end{equation}
Applying the Hankel transform to the Bessel differential equation results in
\begin{equation}
    -\int_0^{\infty}r\Big(\frac{\partial^2\phi}{\partial r^2}+\frac{1}{r}\frac{\partial\phi}{\partial r}-\frac{n^2}{r^2}\phi\Big)J_n(\xi r)\,dr=\xi^2\int_0^{\infty}r\phi(r,z)J_n(\xi r)\,dr\,.
\end{equation}
Replacing in Eq.\ (\ref{eq:Laplace_int}) and using the Hankel transform as described in Eq.\ (\ref{eq:Hankel}), we obtain the ordinary differential equation verified by $\overline{\phi}$:
\begin{equation}
    \mathrm{d}^2\overline{\phi}/\mathrm{d}z^2=\xi^2\overline{\phi}\,.
\end{equation}
The solution is of the form:
\begin{equation}
    \overline{\phi}(\xi,z)=B(a\xi)e^{-\xi z}+C(a\xi)e^{\xi z}\,,
\end{equation}
where the functions $B$ and $C$ are to be determined with the boundary conditions. Applying the inverse Hankel transform defined in Eq.\ (\ref{eq:invHankel}) to this solution allows writing the potential flow $\phi$ as
\begin{equation}\label{eq:potential}
    \phi(r,z)=\int_0^{\infty}\xi[B(a\xi)e^{-\xi z}+C(a\xi)e^{\xi z}]J_n(\xi r)\,d\xi\,.
\end{equation}
The condition on the top casing surface in Eq.\ (\ref{eq:BCcasing}) gives
\begin{equation}\label{eq:linkB-C}
    B(a\xi)e^{-\xi H_{up}}-C(a\xi)e^{\xi H_{up}}=0\,,
\end{equation}
Then isolating $C(a\xi)$ and substituting its expression in Eq.\ (\ref{eq:BCdisk}) yields
\begin{equation}
    \int_0^{\infty}\xi\Big[\xi B(a\xi)\big(1-e^{-2\xi H_{up}}\big)\Big]J_n(\xi r)\,d\xi=-\Big(1+\frac{n\Omega_{D/F}}{\omega}\Big)\psi_n(r)\,.
\end{equation}
Upon using the Hankel transform properties, we obtain
\begin{equation}
    \xi B(a\xi)\big(1-e^{-2\xi H_{up}}\big)=-a\Big(1+\frac{n\Omega_{D/F}}{\omega}\Big)\int_b^a r\,\psi_{ns}(r)J_n(\xi r)\,dr\,.
\end{equation}
This new integral can be evaluated numerically. Isolating $\xi B(a\xi)$ also provides $\xi C(a\xi)$ according to Eq.\ (\ref{eq:linkB-C}). Ultimately adding these two terms leads to
\begin{equation}
    \xi[B(a\xi)+C(a\xi)]=-a\Big(1+\frac{n\Omega_{D/F}}{\omega}\Big)\bigg(\frac{1+e^{-2\xi H_{up}}}{1-e^{-2\xi H_{up}}}\bigg)\int_b^a r\, \psi_{ns}(r)J_n(\xi r)\,dr\,.
\end{equation}
And replacing this expression in Eq.\ (\ref{eq:potential}) gives the expression of the potential flow in the fluid volume:
\begin{align}\label{eq:phiup_vol}
    \phi_{up}(r,z)&=-a^2\omega\Big(1+\frac{n\Omega_{D/F}}{\omega}\Big)\int_0^{\infty}H(a\xi)J_n(\xi r)\bigg[\frac{e^{-\xi z}+e^{\xi(z-2H_{up})}}{1-e^{-2\xi H_{up}}}\bigg]\,d\xi\,,\\\label{eq:phidown_vol}
    \phi_{down}(r,z)&=-a^2\omega\Big(1+\frac{n\Omega_{D/F}}{\omega}\Big)\int_0^{\infty}H(a\xi)J_n(\xi r)\bigg[\frac{e^{-\xi z}+e^{\xi(z-2H_{down})}}{1-e^{-2\xi H_{down}}}\bigg]\,d\xi\,,
\end{align}
where
\begin{equation}
    H(a\xi)=\int_b^a r\,\psi_{ns}(r)J_n(\xi r)\,dr\,.
\end{equation}
The perturbation fluid velocity magnitude depends on the rotation speed and frequency of the considered mode; while the velocity field shape only depends on the modeshape and geometrical properties of the domain. The radial form of the flow potential is determined by the disk modeshape through the imposition of the $z=0$ boundary condition Eq.\ (\ref{eq:BCdisk}). Thus imposing the structural boundary conditions (here clamped-free) also imposes the radial flow conditions. The resulting flow potential Eqs.\ (\ref{eq:phiup_vol}-\ref{eq:phidown_vol}) verifies zero radial speed at the inner radius:
\begin{equation}
    \evalat[\bigg]{\frac{\partial\phi}{\partial r}}{r=b}=0\,.
\end{equation}
However, the radial speed is not zero at the outer radius, and this condition cannot be imposed with this shape of $\phi$. All terms tend towards zero with increasing $r$:
\begin{equation}
    \lim_{r\to\infty}\frac{\partial\phi}{\partial r}=0\,.
\end{equation}
Thus no radial confinement is modelled for $r>a$.

Evaluating the flow potential Eqs.\ (\ref{eq:phiup_vol}-\ref{eq:phidown_vol}) at the disk surface $z=0^+$ yields
\begin{align}\label{eq:phiup}
    \phi_{up}(r,0)&=-a^2\omega\Big(1+\frac{n\Omega_{D/F}}{\omega}\Big)\int_0^{\infty}H(a\xi)J_n(\xi r)G_{up}(a\xi)\,d\xi\,,\\\label{eq:phidown}
    \phi_{down}(r,0)&=-a^2\omega\Big(1+\frac{n\Omega_{D/F}}{\omega}\Big)\int_0^{\infty}H(a\xi)J_n(\xi r)G_{down}(a\xi)\,d\xi\,,
\end{align}
where
\begin{equation}
    G_{up}(a\xi)=\frac{1+e^{-2\xi H_{up}}}{1-e^{-2\xi H_{up}}}\,,\hspace{0.5cm} G_{down}(a\xi)=\frac{1+e^{-2\xi H_{down}}}{1-e^{-2\xi H_{down}}}\,,
\end{equation}
so that $H$ only depends on the mode and on the disk dimensions, and $G_{up},G_{down}$ only depend on the disk and casing dimensions.

In order to assess the influence of the surrounding fluid on the natural frequencies of the structure, we calculate the \ac{AVMI} factor $\beta$, which links natural frequencies in vacuum to natural frequencies in the considered fluid \citep{kwak1991,amabili1996_1}:
\begin{equation}\label{eq:omegbeta}
    \frac{\omega}{\omega_v}=\frac{1}{\sqrt{1+\beta}}\,.
\end{equation}
The \ac{AVMI} factor is expressed as the ratio between the reference kinetic energy of the surrounding fluid $E_F$ to the reference kinetic energy of the disk $E_D$. On the one hand, \citet{lamb1945} provides the expression for $E_F$:
\begin{equation}
    E_F=-\frac{1}{2}\rho_F a\int_0^{2\pi}\int_b^a\frac{\partial\phi}{\partial z}(r,0)\,\phi(r,0)\,r\,dr\,.
\end{equation}
Then substituting Eq.\ (\ref{eq:BCdisk}) and considering both the reference kinetic energies of the above and below fluid:
\begin{equation}
    E_F=-\frac{1}{2}\rho_F a\psi_{\theta}\Big(1+\frac{n\Omega_{D/F}}{\omega}\Big)\int_b^a \big[\phi_{down}(r,0)+\phi_{up}(r,0)\big]\psi_{ns}(r)r\,dr\,,
\end{equation}
where $\psi_{\theta}=2\pi$ if $n=0$ and $\psi_{\theta}=\pi$ otherwise, as results from the integration over $\theta$. Finally replacing the flow potentials by their known expressions \mbox{Eqs.\ (\ref{eq:phiup}-\ref{eq:phidown})}:
\begin{multline}
    E_F=\frac{1}{2}\rho_F a^3\psi_{\theta}\Big(1+\frac{n\Omega_{D/F}}{\omega}\Big)^2\\\times\int_b^a\int_0^{\infty}H(a\xi)J_n(\xi r)\big[G_{down}(a\xi)+G_{up}(a\xi)\big]\,d\xi\,\psi_{ns}(r)r\,dr\,.
\end{multline}
On the other hand, the reference kinetic energy of the disk is
\begin{equation}
    E_D=\frac{1}{2}\rho_D a\psi_{\theta}h\int_b^a \psi_{ns}(r)^2r\,dr\,.
\end{equation}
Therefore, the \ac{AVMI} factor can be expressed as
\begin{multline}\label{eq:beta}
    \beta=\frac{E_F}{E_D}=\rho_Fa\Big(1+\frac{n\Omega_{D/F}}{\omega}\Big)^2\\\times\frac{\int_b^a\int_0^{\infty}aH(a\xi)J_n(\xi r)\big[G_{down}(a\xi)+G_{up}(a\xi)\big]\,d\xi\,\psi_{ns}(r)r\,dr}{\rho_Dh\int_b^a \psi_{ns}(r)^2r\,dr}\,.
\end{multline}
At this point we can see that $\beta$ depends on $\omega$ when the disk is rotating. Let us call $\beta_0$ the expression of the \ac{AVMI} factor when there is no rotation:
\begin{equation}
    \beta_0=\evalat[\big]{\beta\,}{\Omega_{D/F}=0}\,.
\end{equation}
Hence,
\begin{equation}\label{eq:avmi_avmi0}
    \beta=\Big(1+\frac{n\Omega_{D/F}}{\omega}\Big)^2\beta_0\,,
\end{equation}
where $\beta_0$ only depends on the fluid, casing and disk geometry and material parameters, and the $n\Omega_{D/F}=n(1-K)\Omega_D$ term yields the influence of the disk rotation. Both parts depend on the considered mode.

Manipulating the implicit expression given by Eq.\ (\ref{eq:omegbeta}) of the natural frequency as a function of $\beta$, we can write the following explicit formulation for the modal analytical prediction of rotating and submerged disk natural angular frequencies:
\begin{equation}\label{eq:modelanal}
    \omega=\frac{\sqrt{(\beta_0+1)\omega_v^2-\beta_0(n\Omega_{D/F})^2}-n\beta_0\Omega_{D/F}}{\beta_0+1}\,.
\end{equation}
Some integral terms in the developed expression of $\beta_0$ need to be determined numerically. \textsc{Matlab} was chosen to implement the method developed in this section. Results for various sets of geometries and modes can be obtained in only a few seconds.

\subsection{\ac{FSI} numerical model}\label{sec:CFD_method}

In this section we develop a numerical model using results of modal analysis and \ac{CFD}, for the prediction of rotating disks natural frequencies. The model can be applied to arbitrarily complex structures, taking into account the radial gap between the disk and the side walls for instance. The aim for this model is purely to predict natural frequencies using solely \textsc{Ansys CFX}, without the structure coupling module. Such approach is valuable when only a CFX licence is available. Hydraulic turbine efficiency assessment is out of the scope of this study.

\begin{figure}[ht]
    \centering
    \includegraphics[scale=0.55]{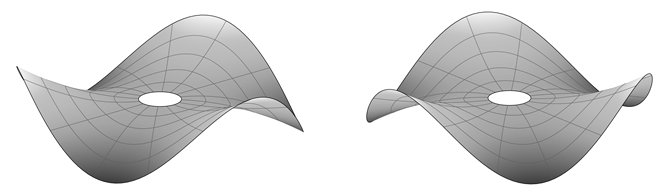}
    \caption{Companion modes $n=3$, $s=0$ for an annular plate}
    \label{fig:compmodes3}
\end{figure}

The oscillating disk in vacuum presents two repeated frequencies for each mode, associated with a co- and counter-rotating wave \citep{ahn1998}. With respect to Eq.\ (\ref{eq:galerkin1}), we need to consider two modes in order to capture both waves. If the disk is rotating in dense fluid, each wave has its own angular frequency, $\omega_+$ and $\omega_-$ respectively. The pair of counter-phased modes $W_c$ and $W_s$ that we consider is called \textit{companion modes}\footnote{$W_s(r,\theta)=W_c(r,\theta-\pi/2n)$} (an example is given in \mbox{Figure \ref{fig:compmodes3}}), where the subscripts $c$ and $s$ refer to the cosine and sine forms of the mode respectively:
\begin{align}
    W_c(r,\theta)=\psi_{ns}(r)\,\cos(n\theta)\,,\\
    W_s(r,\theta)=\psi_{ns}(r)\,\sin(n\theta)\,.
\end{align}
Replacing the solution of the previous section, the vertical displacement is now given by
\begin{equation}\label{eq:dispZ}
    w(r,\theta,t)=W_c(r,\theta)q_c(t)+W_s(r,\theta)q_s(t)\,,
\end{equation}
where $q_c$ and $q_s$ are the two unknowns to be determined by coupling \ac{CFD} with a mass-spring system, namely the two generalized coordinates of the 2DOF problem. Because the mass-spring Eq.\ (\ref{eq:1dof}) solved by $w$ is linear, it is equivalent to the following system solved by $q_c$ and $q_s$:
\begin{align}\label{eq:maxZc}
    \ddot{q_c}(t)+\omega^2q_c(t)=\frac{F_c(t)}{M} &\,, \text{ where } F_c(t)=\int_{r=b}^a\int_{\theta=0}^{2\pi}P(r,\theta,t)W_c(r,\theta)\,rdrd\theta\,,\\
    \ddot{q_s}(t)+\omega^2q_s(t)=\frac{F_s(t)}{M} &\,, \text{ where } F_s(t)=\int_{r=b}^a\int_{\theta=0}^{2\pi}P(r,\theta,t)W_s(r,\theta)\,rdrd\theta\,.\label{eq:maxZs}
\end{align}
The modal forces $F_c$ and $F_s$ are the projections of the pressure fields on the respective modeshapes $W_c$ and $W_s$. The angular frequency $\omega$ and the modal mass $M$ are unchanged because they only depend on the structural properties. Water effects are taken into account through the modal force. Solving these equations with \ac{CFD} provides time signals for $q_c(t)$ and $q_s(t)$, from which the angular frequencies $\omega_+$ and $\omega_-$ can be deduced. It should be noted that whereas the structural equations are linear,the \ac{CFD} calculations are not, making the coupled system nonlinear.

\begin{figure}[t]
    \centering
    \includegraphics[scale=0.75]{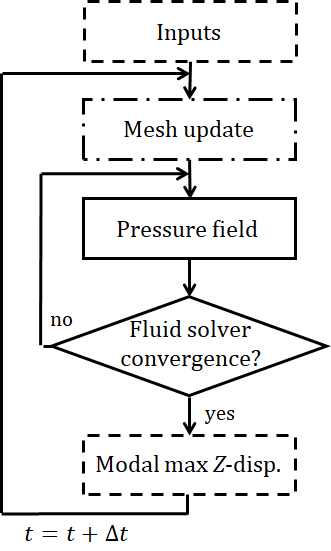}
    \caption{Working scheme for the numerical model. Continuous boxes symbolize steps solved within \textsc{CFX}, while dashed boxes represent steps solved with external user \textsc{Fortran} subroutines (Junction Box). The mesh update consists of solving Eq.\ (\ref{eq:dispZ}). The maximum \mbox{$Z$-displacement} step is achieved with the resolution of Eqs.\ (\ref{eq:maxZc}-\ref{eq:maxZs}) using an adapted Runge-Kutta algorithm. Convergence is based on RMS criteria of conservative control volume fluid equation residuals.}
    \label{fig:NumScheme}
\end{figure}

\begin{figure}[t]
    \centering
    \includegraphics[width=\textwidth]{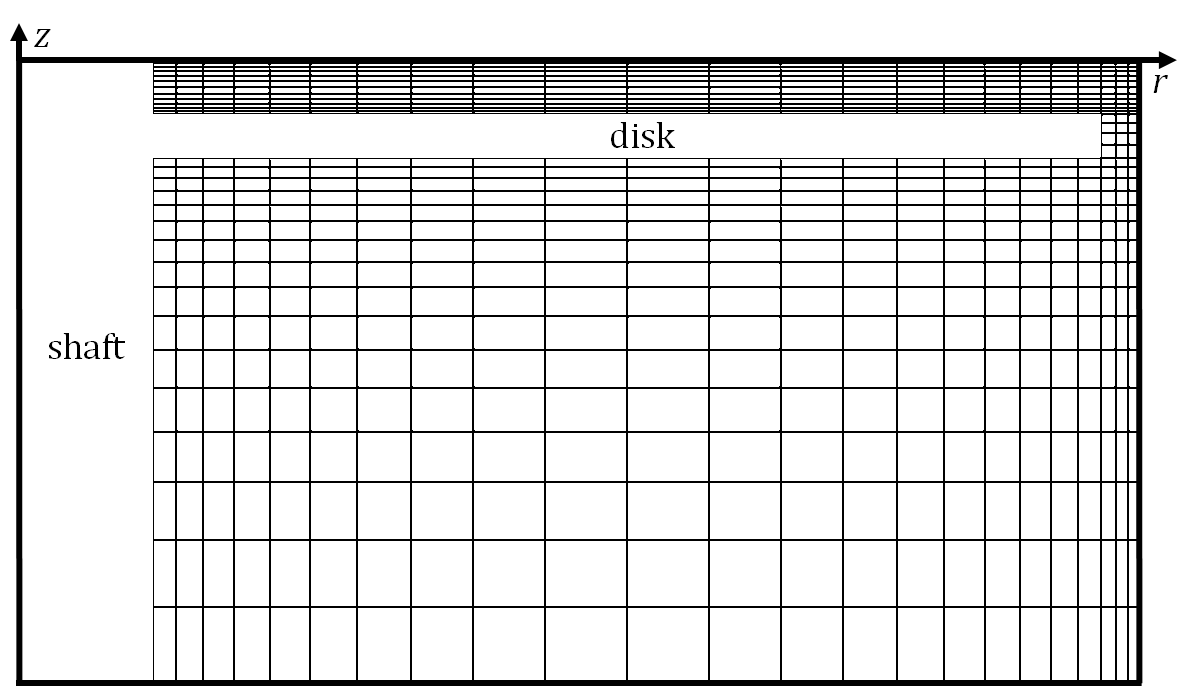}
    \caption{Fluid domain mesh of the \citet{presas2015} experimental test rig in the ($r,z$) plan for the \ac{CFD} model. Elements are cubic hexagonal, regularly spaced in the $\theta$ direction at intervals of 2$^\circ$. There are $118080/n$ elements in this mesh, where $n$ is the number of nodal diameters of the studied mode.}
    \label{fig:Mesh}
\end{figure}

The \ac{CFD} method developed in this section was implemented with \textsc{Ansys CFX} 18.2. The fluid equations are solved by \textsc{Ansys CFX} itself, while the disk displacement according to the imposed modeshape is implemented with custom user CEL functions and \textsc{Fortran} routines (see \mbox{Figure \ref{fig:NumScheme}}). \textsc{Ansys CFX} performs \ac{URANS} calculations with a second order backward Euler transient scheme. We use CFX \texttt{High resolution} advection scheme, which means the code tries second order where possible and reduces to first order where convergence is compromised. More details on CFX model implementation can be found in \textsc{Ansys CFX} User's Manual. For a mode with $n$ nodal diameters, only a fraction $1/n$ of the actual geometry is represented and periodic boundary conditions are applied consequently. The fluid domain mesh (presented in Figure \ref{fig:Mesh}) is composed of approximately \mbox{$10^4$--$10^5$} cubic hexagonal elements, depending on the studied mode. As the fluid rotates with the disk, their are higher gradients for smaller axial gaps, hence the finer mesh above the disk. This coarse mesh is sufficient for our simulations because mode split is an inviscid fluid phenomenon, hence it is not influenced by viscous effects in the boundary layer. Mesh convergence was ensured, calculating a 1.96\% frequency relative error with finer meshes (refinement factor of 2 in all three dimensions). The radial gap is considered in this model.

The model first requires input parameters. The cosine and sine form of the modeshape are established using Eq.\ (\ref{eq:modeshape}), and normalized according to \mbox{Eq.\ (\ref{eq:norm})}. The Bessel functions are approximated with polynomials. The modal mass and rigidity of the structure are determined from our structural analysis in \S \ref{sec:struct}, using the disk material properties: \mbox{$M=\rho_Dh$} and \mbox{$K_{ns}=Dk_{ns}^4=M\omega_v^2$}. The surrounding dense fluid, typically water for hydraulic turbine applications, is considered compressible to avoid numerical instability issues due to pressure wave propagation, as indicated in the guidelines provided by \textsc{Ansys}. Preliminary simulations showed no difference in the predicted frequencies between the $k$-$\omega$ SST and $k$-$\epsilon$ turbulence models. All simulations performed for this paper used the latter for its better numerical robustness. The mesh respects $Y^+>30$ in accordance with the turbulence model. The time step duration $\Delta t$ is paramount to achieve numerical stability in this case of high fluid-solid density ratio \citep{wong2013}. Typically, for steel runners in water, $\rho_F/\rho_D\approx0.1$. Depending on the studied geometry, the choice of $\Delta t$ may be critical to stabilize the calculation. $\Delta t\sim10^{-6}$ s in our work, and $t_0=\sqrt{\rho_Dh/D}=8.17\cdot10^{-2}$ s for the disk of the \citet{presas2015} test rig. Hence, the dimensionless time $\Delta\tau=\Delta t/t_0\sim10^{-5}$ grants enough precision so that the fluid equations time discretization scheme needs not be of high order. Moreover, the highest mesh displacement in one time step is less than a hundred times the smallest cell size.

We then initialize the model in the rotating disk reference frame by setting the domain angular velocity to $\Omega_D$. The rotating parts in the stationary reference frame are therefore considered stationary in the rotating disk reference frame. Casing walls are defined as counter-rotating. Centrifugal and Coriolis forces are accounted for in the fluid. At first, performing a steady state computation with $w=0$ imposed allows the flow to stabilize in the rotating domain. Then enabling the mesh vertical motion and applying a sine pulse to the modal force during the first steps induces movement. We then leave the system to oscillate freely.

During the main calculation, we apply the following procedure to each time step of the simulation, as shown in Figure \ref{fig:NumScheme}:
\begin{enumerate}
    \item Pressure and velocity fields are computed in the fluid domain.
    \item The pressure field is integrated on the disk, for each modeshape. The second order \mbox{Eq.\ (\ref{eq:1dof})} is converted to a first order system of equations and then solved using the Runge-Kutta algorithm.
    \item The new total mesh $Z$-displacement is calculated according to Eq.\ (\ref{eq:dispZ}). The mesh is then updated.
\end{enumerate}
A frequency analysis of the $Z$-displacement time signal provides the free oscillation frequencies of the system for the chosen mode.

\section{Results and Discussion}

\subsection{Analytical model results}

\begin{figure}[t]
    \centering
    \includegraphics[scale=0.55]{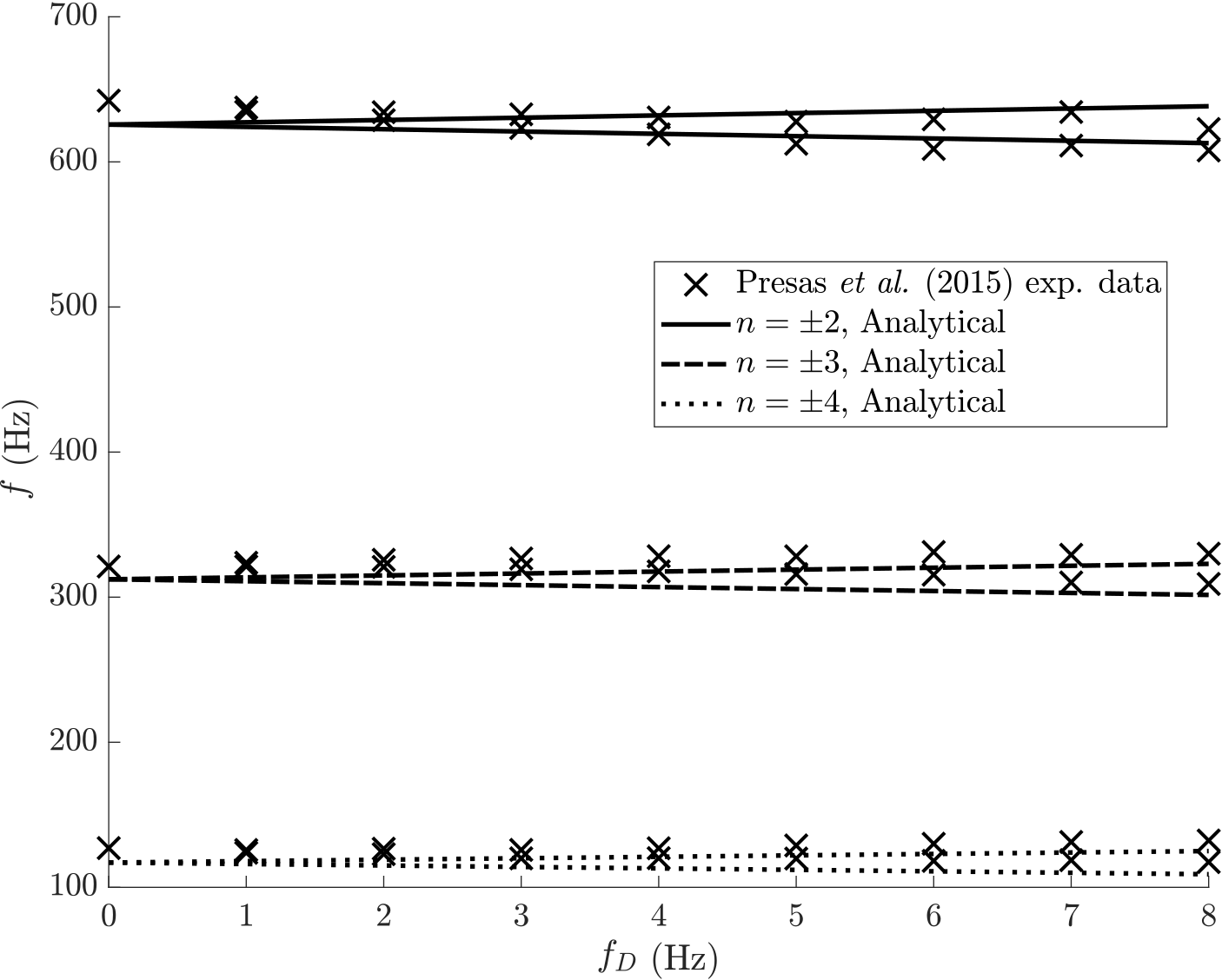}
    \caption{Comparison of the disk natural frequencies for modes $n=\pm\,2,3,4$, $s=0$ and the \citet{presas2015} test rig geometry; dotted data corresponds to their experimental results and lines where obtained with the analytical model detailed in this section. $f=|\omega|/2\pi$.}
    \label{fig:expPres-anal}
\end{figure}

In this section we analyse the modal analytical model for rotating and submerged disk natural frequency prediction, given by Eq.\ (\ref{eq:modelanal}). Table \ref{tab:parameters} summarizes the parameters used to model the \citet{presas2015} experimental test rig. Table \ref{tab:Pres-anal} and \mbox{Figure \ref{fig:expPres-anal}} present frequencies of this structure's natural frequencies for modes $n=2,3,4$ nodal diameters and $s=0$ nodal circles, assessed with \mbox{Eq.\ (\ref{eq:modelanal})}. Their test rig uses a rotating disk in water, confined in a rigid casing, with small radial gap, which makes it particularly relevant with regards to our analytical model hypotheses. For a given mode, the natural frequencies of both co- and counter-rotating waves are identical when the disk is stationary. The split between $\omega_+$ and $\omega_-$ then increases with the rotation speed, while the average value slightly decreases. Overall, Figure \ref{fig:expPres-anal} illustrates that the model shows good accuracy with respect to \citet{presas2015} experimental results.

\begin{table}[ht]
    \centering
    \caption{Parameter values for modeling the \citet{presas2015} experimental test rig geometry. The disk is made of stainless steel; the fluid is water.}
    \begin{tabular}{c c@{\hskip 2 cm}c c}
        $E$ & 200$\cdot10^9$ Pa & $\nu$ & 0.27 \\
        $\rho_D$ & 7680 kg/m$^3$ & $\rho_F$ & 997 kg/m$^3$\\
        $a$ & 0.2 m & $b$ & 0.025 m\\
        $H_{up}$ & 0.01 m & $H_{down}$ & 0.097 m\\
        $h$ & 0.008 m & $K$ & 0.45
    \end{tabular}
    \label{tab:parameters}
\end{table}
\begin{table}[ht]
    \centering
    \caption{Natural frequencies of modes $n=\pm\,2,3,4$, $s=0$ obtained with the analytical model Eq.\ (\ref{eq:modelanal}) and the \citet{presas2015} experiments for different disk rotation speeds and the corresponding test rig geometry, and relative error $\epsilon$. $f=|\omega|/2\pi$.}
    \begin{tabular}{c|c c c|c c c|c c c}
        [Hz] & \multicolumn{3}{c|}{$n=2$} & \multicolumn{3}{c|}{$n=3$} & \multicolumn{3}{c}{$n=4$}\\
        \hline
        $f_D$ & 0 & 4 & 8 & 0 & 4 & 8 & 0 & 4 & 8\\
        $f_{+,\rm{exp}}$ & 127.1 & 120.1 & 117.4 & 321.2 & 317.8 & 309.1 & 642.2 & 619 & 607.9\\
        $f_{+,\rm{ana}}$ & 117.1 & 113.4 & 109.7 & 312.2 & 307.3 & 302.4 & 626.1 & 620.3 & 614.4\\
        $\epsilon_+$ & 7.9\% & 5.6\% & 6.6\% & 2.8\% & 3.3\% & 2.2\% & 2.5\% & 0.2\% & 1.1\%\\
        $f_{-,\rm{exp}}$ & 127.1 & 126.9 & 132.3 & 321.2 & 328.2 & 330.0 & 642.2 & 630.6 & 633.7\\
        $f_{-,\rm{ana}}$ & 117.1 & 120.7 & 124.4 & 312.2 & 317.1 & 322.0 & 626.1 & 631.9 & 637.7\\
        $\epsilon_-$ & 7.9\% & 7.7\% & 4.9\% & 2.8\% & 4.9\% & 2.4\% & 2.5\% & 0.2\% & 0.6\%\\
    \end{tabular}
    \label{tab:Pres-anal}
\end{table}

\begin{figure}[t]
    \centering
    \includegraphics[scale=0.55]{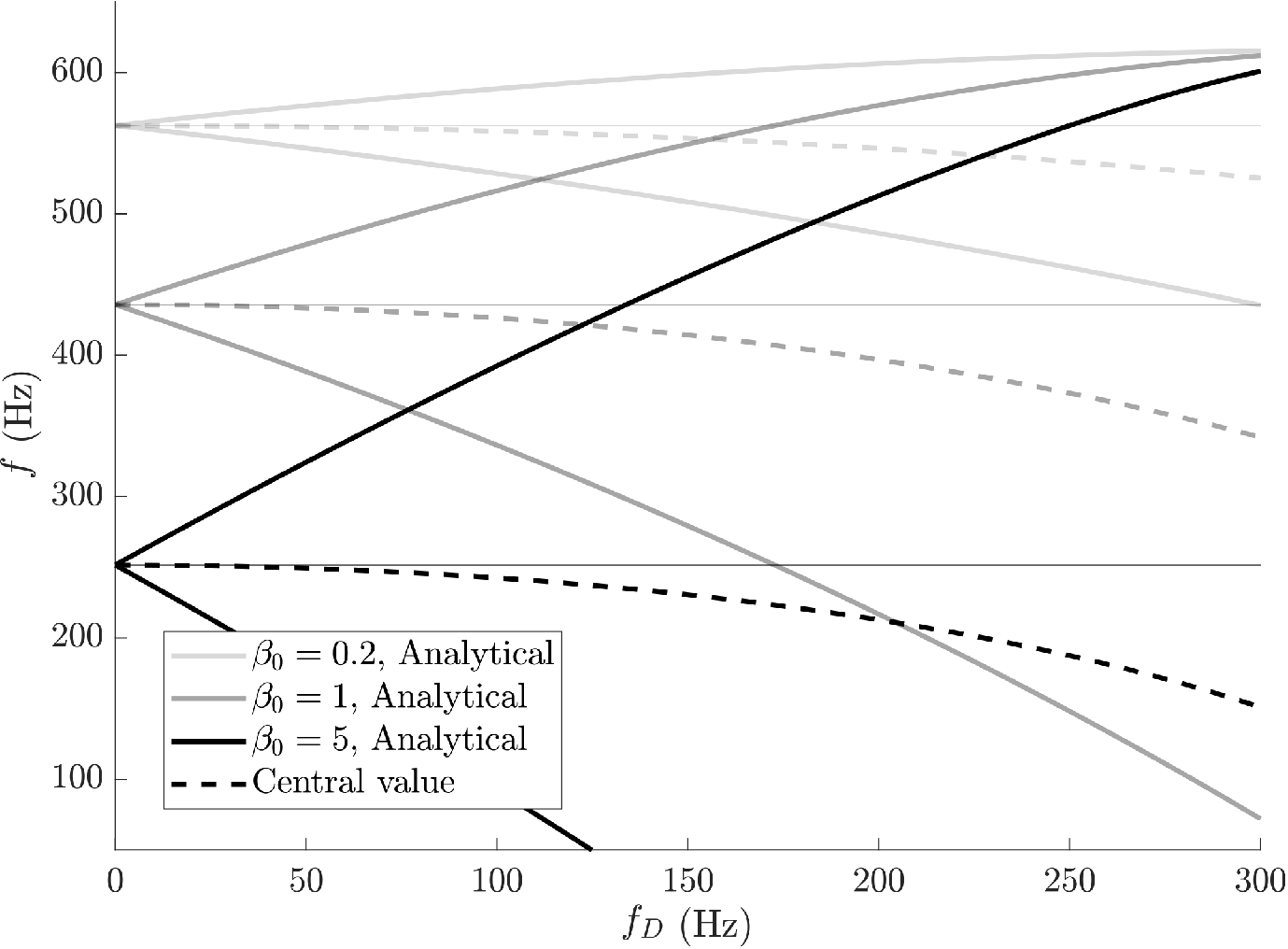}
    \caption{Disk analytical frequencies for mode $n=\pm3$, $s=0$ and a large range of disk rotating speeds. The system geometry is the same for all curves, but $\beta_0$ varies \mbox{from 0.2 to 5}. Black curves represent co- and counter-rotating mode frequencies, and the black dotted line shows the deviation from the $\Omega_D=0$ natural frequency. This deviation is largest for $\beta_0=1$, while the mode split magnitude increases with $\beta_0$. Hydraulic turbines typically have \mbox{$\Omega_D/2\pi\leq10$ Hz}. $f=|\omega|/2\pi$.}
    \label{fig:beta0-anal}
\end{figure}

Our predictive analytical equation provides information on the two physical phenomena applying to disks rotating in a dense fluid. First, let us recall that $n>0$ and $n<0$ respectively characterize co- and counter-rotating waves relative to the rotating disk. Therefore, the $-n\beta_0\Omega_{D/F}$ term in Eq.\ (\ref{eq:modelanal}) is responsible for the mode split phenomenon: the natural frequency is increased by the disk rotation for counter-rotating waves, while it is decreased for co-rotating waves. The mode split magnitude is then given by
\begin{equation}\label{eq:split_amp}
    \omega_--\omega_+=\frac{2n\beta_0\Omega_{D/F}}{\beta_0+1}\,.
\end{equation}
Secondly, the $-\beta_0(n\Omega_{D/F})^2$ term inside the square root of Eq.\ (\ref{eq:modelanal}) is responsible for the frequency drift: it decreases the value of the frequency regardless of the propagation direction of the wave. Both of these terms are proportional to the relative rotation speed between the disk and the fluid multiplied by the number of nodal diameters. However, the frequency drift magnitude is smaller than the mode split magnitude for typical rotation speeds of high head runners. Another way to interpret the analytical results is to rearrange Eq.\ (\ref{eq:avmi_avmi0}) into
\begin{equation}
    \frac{\beta-\beta_0}{\beta_0}=\frac{2n\Omega_{D/F}}{\omega}+\frac{n^2\Omega_{D/F}^2}{\omega^2}=2\widetilde{U}+\widetilde{U}^2\,,
\end{equation}
where $\widetilde{U}=n\Omega_{D/F}/\omega$ is the relative circumferential wave speed. The influence of the rotation on the \ac{AVMI} factor then depends linearly and quadratically on the relative circumferential wave speed. The linear term causes the frequency split, whereas the quadratic term leads to the frequency drift. In conclusion, mode split and drift respectively have their physical origin in linear and quadratic dependence of the added mass with the relative wave speed.

The influence of the considered mode and of both the disk and casing geometries on the structure natural frequencies is more difficult to interpret because of the complexity of the $\beta_0$ parameter. A parametric study shows that if the axial gap is large enough ($H_{up},H_{down}\geq0.2\,a$), $\beta_0$ is in the order of magnitude of the density ratio multiplied by the aspect ratio of the disk, while it tends towards infinity if this gap becomes null:
\begin{equation}\label{eq:beta0asympt}
    \evalat[\Big]{\beta_0}{H_{up},H_{down}\geq0.2\,a}\propto\frac{\rho_F}{\rho_D}\frac{a}{h}\,,\hspace{1cm}\lim_{H\to0}\beta_0=\infty\,.
\end{equation}
Independently, it can be determined from Eq.\ (\ref{eq:beta}) that $\beta_0$ increases with shorter axial gaps, and with larger and thinner disks. Additional numerical tests show that $\beta_0$ decreases with the number of nodal diameters and circles. Figure \ref{fig:beta0-anal} presents the influence of $\beta_0$ on the natural frequencies predicted with our analytical model. The mode split magnitude increases with $\beta_0$ towards the asymptotic value of $2n\Omega_{D/F}$, which is also apparent from Eq.\ (\ref{eq:split_amp}). The frequency drift magnitude presents a maximum for $\beta_0=1$. For typical turbine runner dimensions, rotation speeds ($\Omega_D/2\pi\leq10$ Hz) and water density, the drift represents less than 1\% of the predicted frequency. In conclusion, the frequency drift effect is negligible in terms of hydraulic turbine applications. This is not true for the frequency split, which needs to be predicted accurately.

\begin{figure}[t]
    \centering
    \includegraphics[scale=0.55]{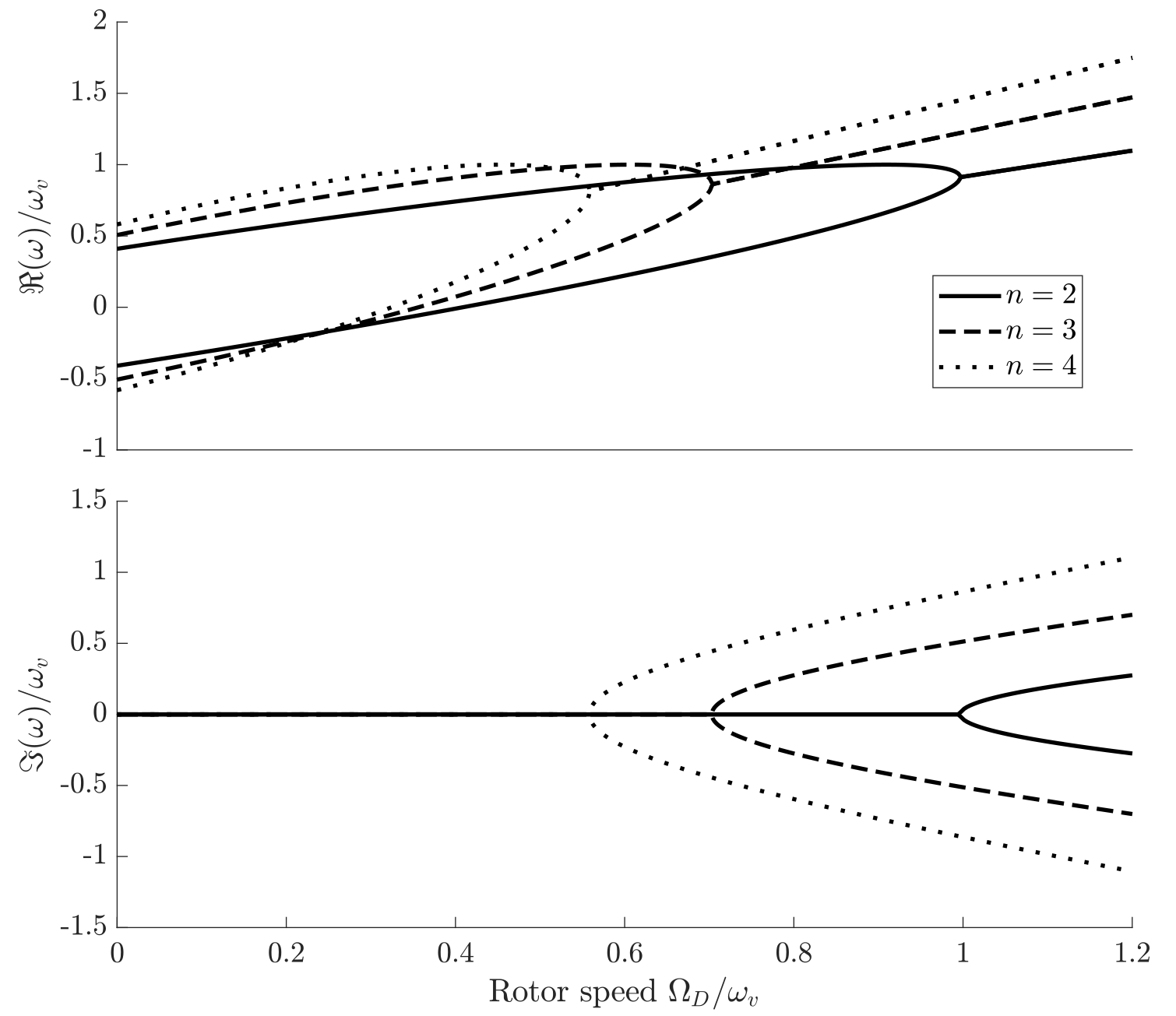}
    \caption{Real (top) and imaginary (bottom) parts of the \citet{presas2015} disk natural frequencies for modes $n=2,3,4$, $s=0$. The two real natural frequencies of a single rotating mode eventually merge when the rotation speed reaches the critical speed. According to \mbox{Eq.\ (\ref{eq:critspeed})}, the corresponding critical speeds are $\Omega_{C,n=2}/2\pi=157$ Hz, $\Omega_{C,n=3}/2\pi=238$ Hz and $\Omega_{C,n=4}/2\pi=331$ Hz.}
    \label{fig:omegflutter}
\end{figure}

\begin{figure}[t]
    \centering
    \includegraphics[scale=0.6]{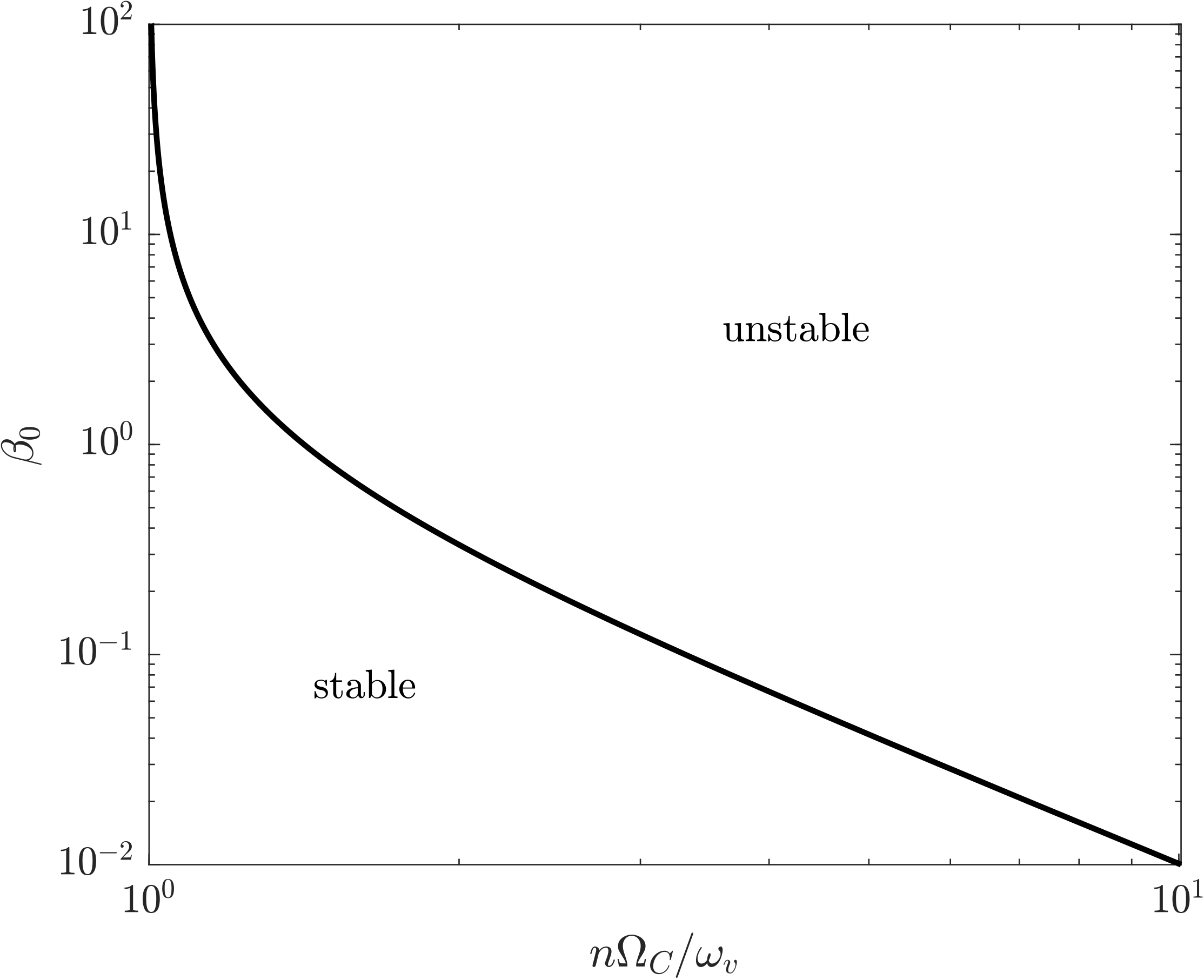}
    \caption{Log-log stability map for the \citet{presas2015} experimental test rig geometry. For a given value of $\beta_0$, any rotation speed corresponding to a point on the right of the line is associated with coupled-mode flutter. The shape of the boundary is similar for any disk geometry, and given by Eq.\ (\ref{eq:critspeed}).}
    \label{fig:stab_map}
\end{figure}

We now consider both positive and negative frequencies (hence $n>0$). Figure \ref{fig:omegflutter} presents the evolution of these natural frequencies with the rotation speed for several modes of the \citet{presas2015} experimental test rig geometry. The frequencies become complex values for $\Omega_{D/F}$ above a critical value $\Omega_C$. This happens when the term under the square root of Eq.\ (\ref{eq:modelanal}) becomes negative, which triggers an unstable coupling between the disk movements and the variation of pressure in the surrounding fluid \citep{kornecki1978,kang2004}. The associated instability is flutter \citep{huang1995,kim2000}, representing a classical Hopf bifurcation, characterized by pairs of natural frequencies with a non-zero imaginary part for $\Omega_{D/F}>\Omega_C$ \citep{paidoussis1998}. The critical disk to fluid rotation speed $\Omega_C$ is given by
\begin{equation}\label{eq:critspeed}
    \frac{n\Omega_C}{\omega_v}=\sqrt{1+\frac{1}{\beta_0}}\,.
\end{equation}

Figure \ref{fig:stab_map} shows the stability map for the non-dimensional critical speed as a function of $\beta_0$. Two asymptotic regimes emerge:
\begin{equation}\label{eq:OmegaRasympt}
    \lim_{\beta_0\to0}\Omega_C=\infty\,,\hspace{1cm}\lim_{\beta_0\to\infty}\Omega_C=\omega_v/n\,.
\end{equation}
This second part in Eq.\ (\ref{eq:OmegaRasympt}) shows that critical speeds are beyond the hydraulic turbine applications range for any mode and typical geometries.

\subsection{Numerical model results}\label{sec:numres}

\begin{figure}[t]
    \centering
    \includegraphics[scale=0.6]{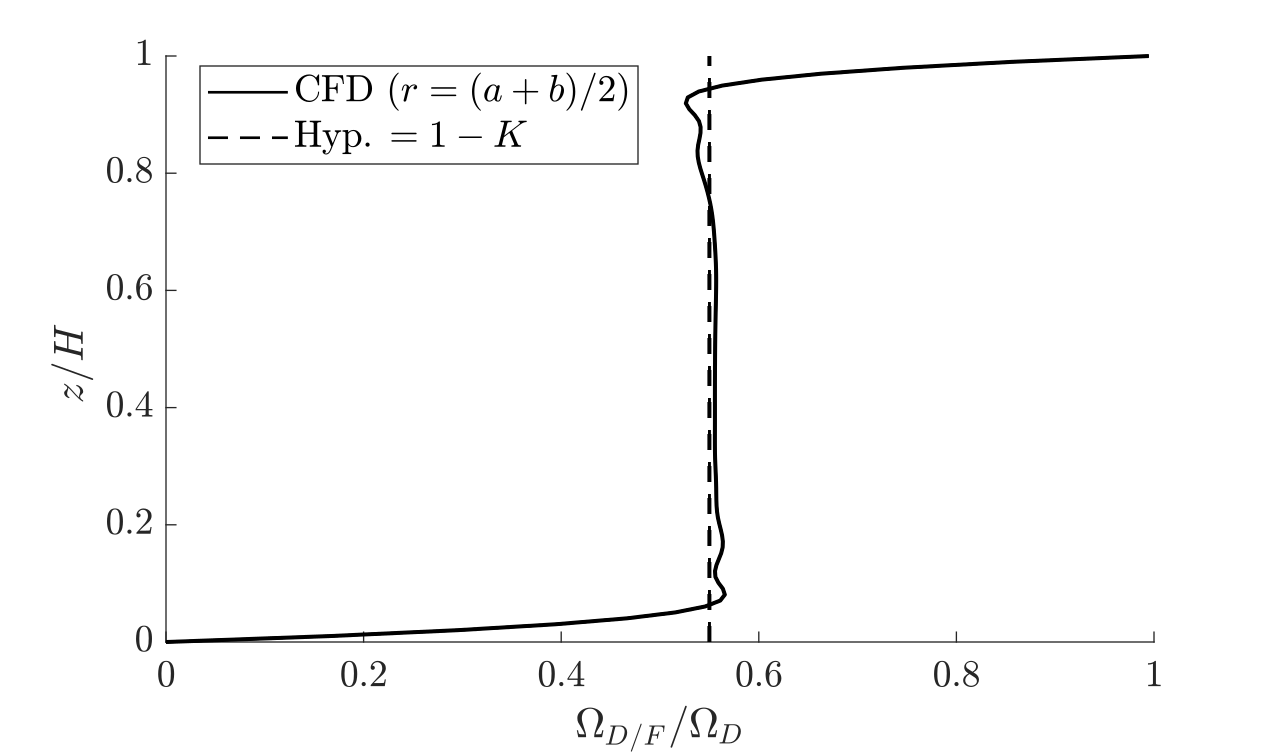}
    \caption{Rotation speed of the disk relative to the fluid in the axial gap above the disk at the middle radius $r=(a+b)/2$. Results were obtained with \ac{CFD} for the \citet{presas2015} experimental test rig geometry rotating at $\Omega_D/2\pi=4$ Hz; $z=0$ corresponds to the rotor surface, while $z=H_{up}$ corresponds to the top part of the casing. The values agree with the theoretical expression of $\Omega_{D/F}/\Omega_D=1-K$ and $K=0.45$ (dashed line).}
    \label{fig:omegaFD_z}
\end{figure}

In this section we validate the hypothesis made on the entrainment coefficient $K$, and we analyze the modal numerical model results for stationary or rotating disks in air or water. All geometrical and physical parameters remain as given in Table \ref{tab:parameters}, and the radial gap is 7 mm long. According to \citet{poncet2005}, the average entrainment coefficient of the flow between a radially delimited rotating and stationary frame satisfies
\begin{equation}
    K\approx0.45\ \textrm{for}\ 10^6<Re=\Omega_Da^2/\nu<4.5\cdot10^6\,,
\end{equation}
where $Re$ is the Reynolds number and $\nu$ is the kinematic viscosity of the fluid. Figure \ref{fig:omegaFD_z} presents the fluid rotation speed relative to the disk in the top axial gap, computed with \ac{CFD}. The fluid obeys the no-slip boundary condition and rotates with the disk on its surface ($\Omega_{D/F}=0$ at $z=0$), it is stationary on the casing surface ($\Omega_{D/F}=\Omega_D$ at $z=H_{up}$), and it rotates at $\Omega_{D/F}\approx0.55\,\Omega_D$ for $0.1<z/H_{up}<0.9$. This verifies that $K\approx0.45$ for the tested geometry, and validates the hypothesis made in the analytical model development.

\begin{figure}[t]
    \centering
    \includegraphics[scale=0.6]{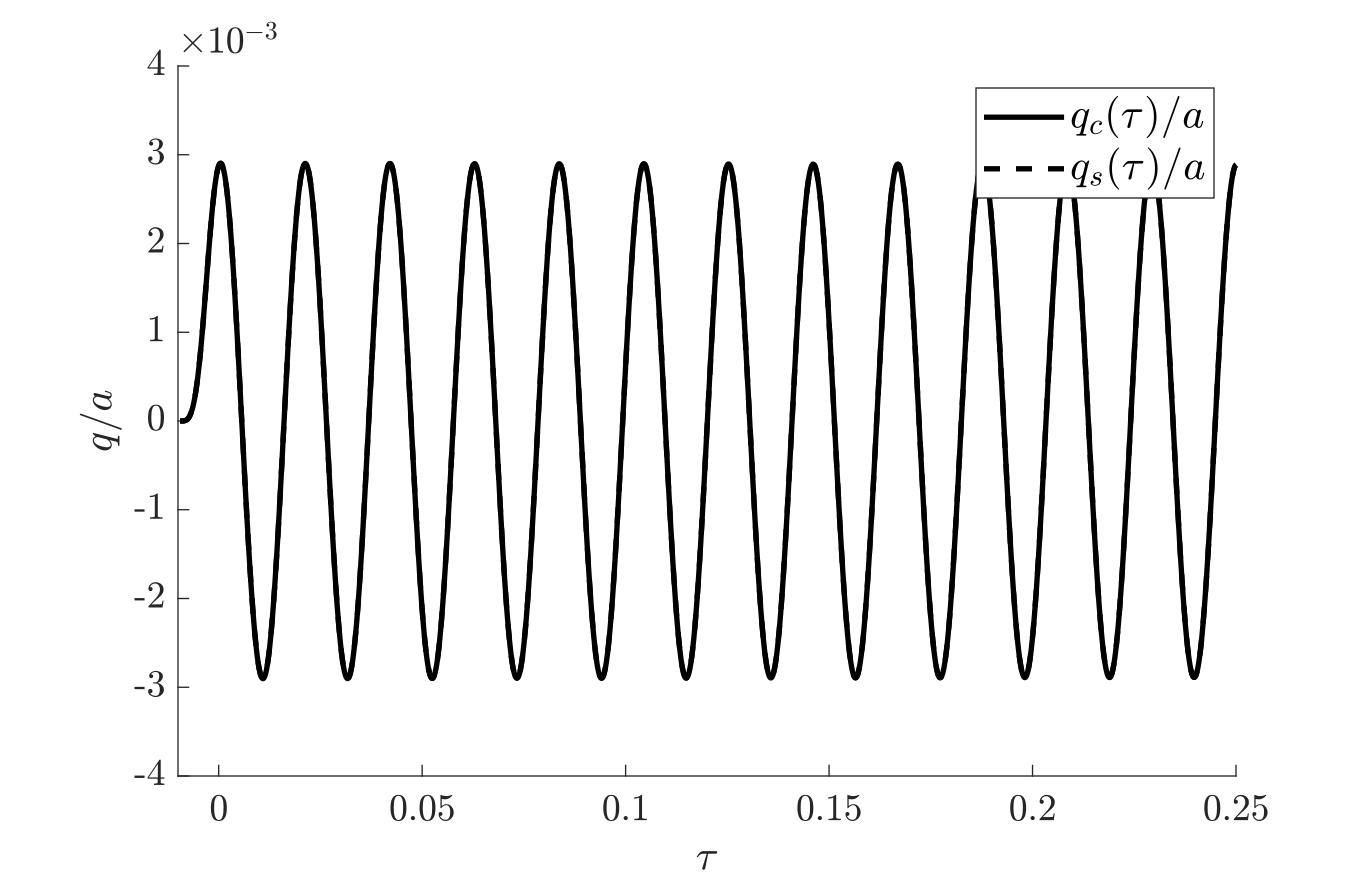}
    \caption{$q_c/a$ and $q_s/a$ as a function of the elapsed dimensionless time for mode $n=3$, $s=0$ and the \citet{presas2015} geometry. The simulation begins with an initial sine pulse, followed by free oscillations of the standing disk in air. Both signals have identical frequencies and the fluid damping is negligible. Structural damping is not taken into account in the \ac{CFD} analysis. Here $\omega/2\pi=616.2$ Hz matches the structure natural frequency.}
    \label{fig:UcUs_air}
\end{figure}

We introduce the non-dimensional time $\tau=t/t_0$, where $t_0=\sqrt{\rho_Dh/D}$ is a reference time. Figure \ref{fig:UcUs_air} presents the $q_c$ and $q_s$ signals from \mbox{Eqs.\ (\ref{eq:maxZc}-\ref{eq:maxZs})} for a disk in air. We matched the disk geometry with the \citet{presas2015} experimental test rig. The vertical displacement appears as periodic and harmonic, with angular frequency \mbox{$\omega_+=\omega_-$}. With low density fluids such as air at 25$^{\circ}$C, the influence of the fluid and of the disk rotation on the natural frequencies is negligible, as Eq.\ (\ref{eq:beta0asympt}) implies \mbox{$\beta_0\sim10^{-3}\ll1$} for geometries of interest. Hence, the identical frequencies for both signals. The fluid damping is also very low.

\begin{figure}[t]
    \centering
    \includegraphics[scale=0.6]{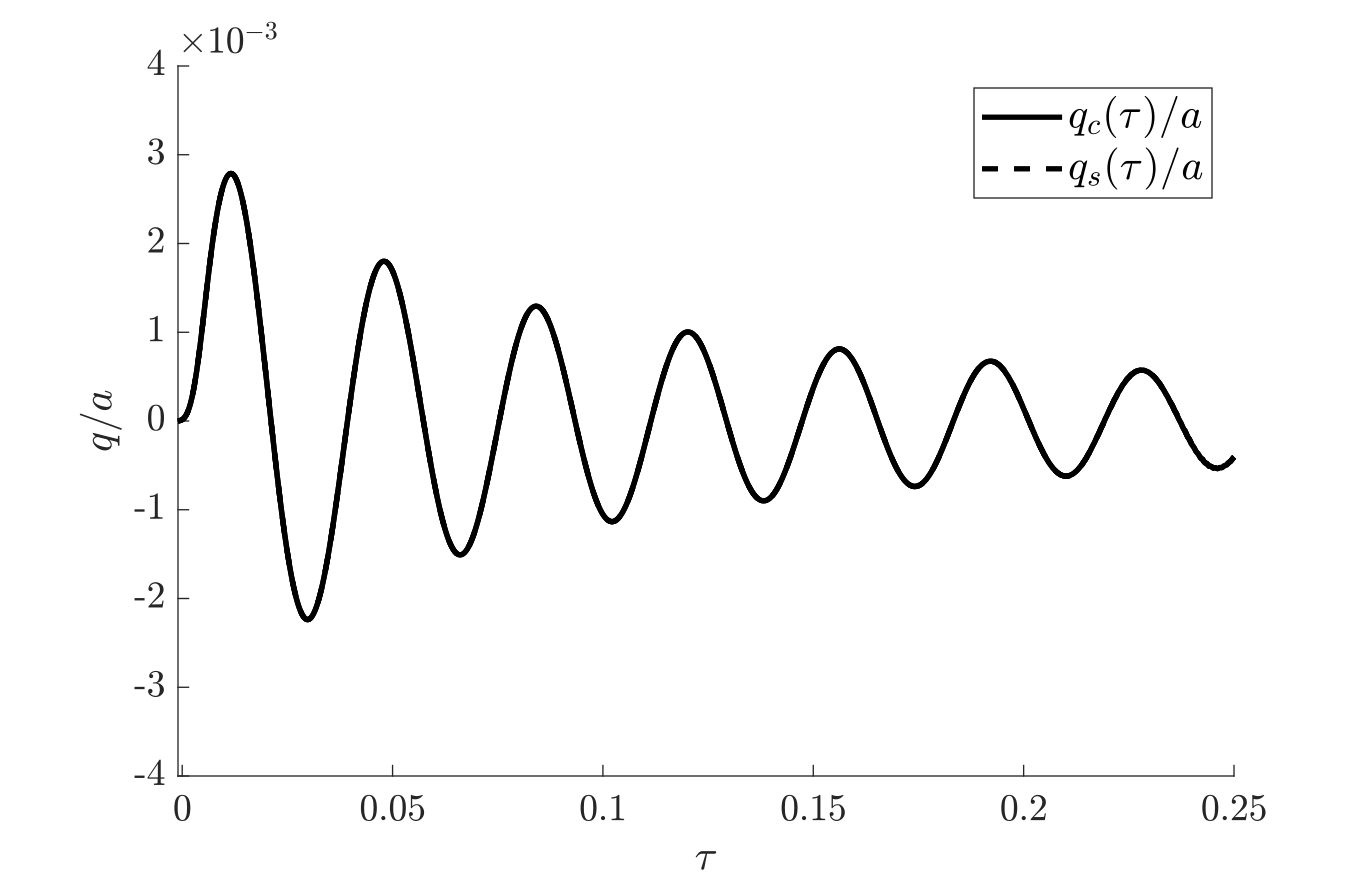}
    \caption{$q_c/a$ and $q_s/a$ as a function of the elapsed dimensionless time for mode $n=3$, $s=0$ and the \citet{presas2015} geometry. The simulation begins with an initial sine pulse, followed by free oscillations of the standing disk in water. Both signals have identical frequencies and their amplitude is damped by the dense fluid. Here $\omega/2\pi=340.3$ Hz agrees with both the analytical model and the experimental data.}
    \label{fig:UcUs_stat_water}
\end{figure}

Figure \ref{fig:UcUs_stat_water} presents the $q_c$ and $q_s$ signals for the same stationary disk, in water. With no rotation, we still have $\omega_+=\omega_-$. However, the signal amplitude now decreases with time. Computed frequencies for several modes are compared, and agree, with the \citet{presas2015} experimental results and with Eq.\ (\ref{eq:modelanal}) in Table \ref{tab:stat_water}. The presence of the radial gap and radial confinement in the \ac{CFD} model is thought to be largely responsible for the small differences between analytical and numerical results. With high density fluids such as water, the fluid influence is no longer negligible, as Eq.\ (\ref{eq:beta0asympt}) implies $\beta_0\sim1$ for geometries of interest. For non-rotating disks, the two main effects of the fluid are the decrease of structural natural frequencies and the damping of the displacement amplitude.
\begin{table}[ht]
    \centering
    \caption{Natural frequencies of modes $n=2,3,4$, $s=0$ obtained with the analytical model Eq.\ (\ref{eq:modelanal}), the numerical model Eq.\ (\ref{eq:1dof}), and experiments from \citet{presas2015} for the stationary disk in water, and relative error $\epsilon$. $f=|\omega|/2\pi$.}
    \begin{tabular}{c|c|c|c}
        [Hz] & $n=2$ & $n=3$ & $n=4$\\
        \hline
        $f_{\rm{exp}}$ & 127.1 & 321.2 & 642.2\\
        $f_{\rm{ana}}$ & 117.1 & 312.2 & 626.1\\
        $\epsilon_{\rm{ana-exp}}$ & 7.9\% & 2.8\% & 2.5\%\\
        $f_{\rm{num}}$ & 130.1 & 340.3 & 641.2\\
        $\epsilon_{\rm{num-exp}}$ & 2.4\% & 5.9\% & 0.2\%
    \end{tabular}
    \label{tab:stat_water}
\end{table}

\begin{figure}[t]
    \centering
    \includegraphics[scale=0.6]{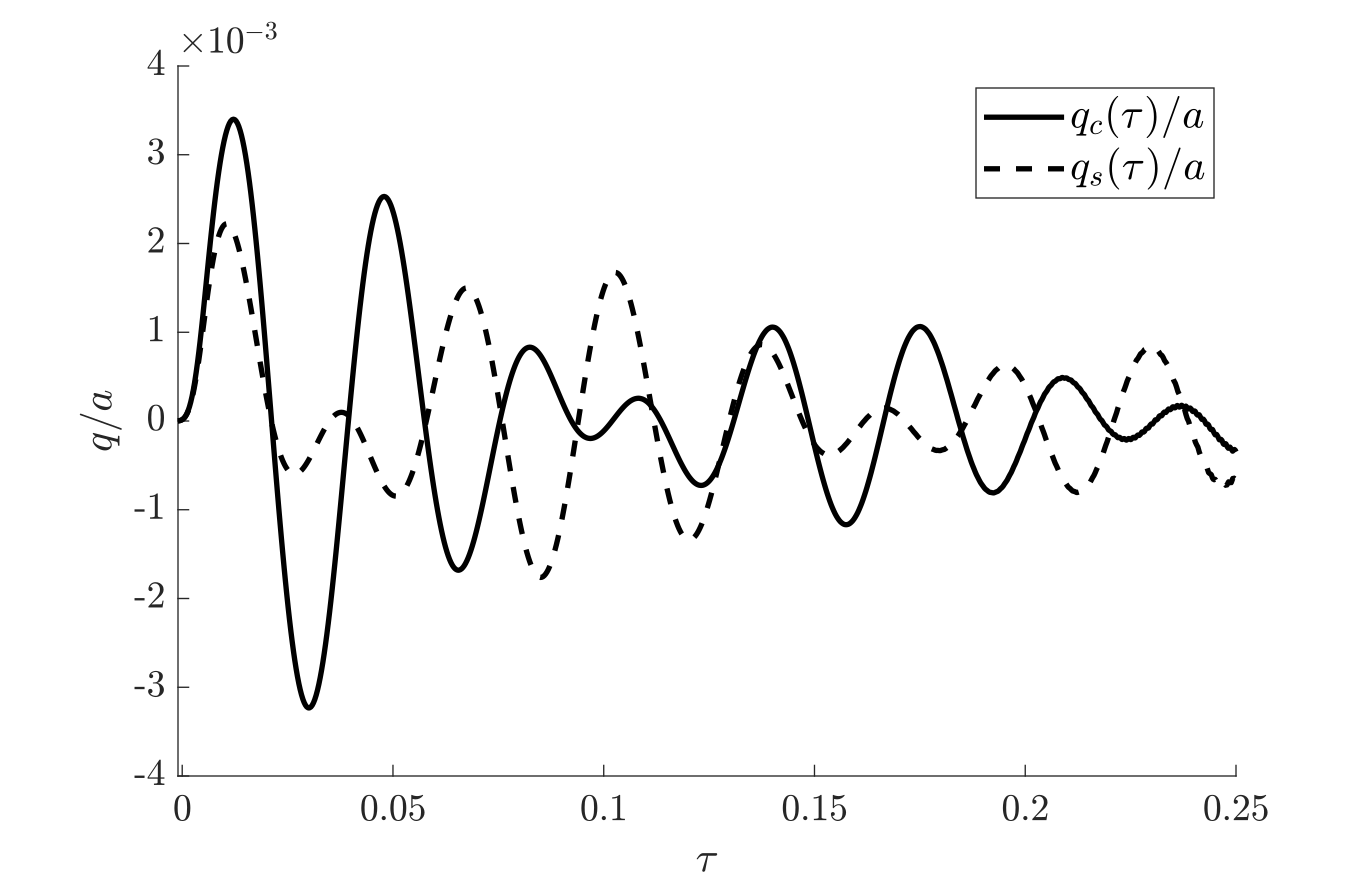}
    \caption{$q_c/a$ and $q_s/a$ as a function of the elapsed dimensionless time for mode $n=3$, $s=0$ and the \citet{presas2015} geometry. The simulation begins with an initial sine pulse, followed by free oscillations of the rotating disk in water ($\Omega_D/2\pi=4$ Hz). Both signals have close but different frequencies, which results in a beating oscillation, characteristic of free vibration under mode split. Here $|\omega_--\omega_+|/2\pi=95.9$ Hz agrees with the analytical model.}
    \label{fig:UcUs_rot_water}
\end{figure}

Figure \ref{fig:UcUs_rot_water} presents the $q_c$ and $q_s$ signals for the same disk, rotating in water. The signal is still damped because of the dense fluid. Adding the rotation of the disk in dense fluid triggers the mode split phenomenon, because of the fluid's different interaction with the co- and counter-rotating waves, as shown in \mbox{Eq.\ (\ref{eq:modelanal})}. Both signals now show a beat envelope, due to the presence of two close frequencies: $\omega_+$ and $\omega_-$. The high frequency can be interpreted as the natural angular frequencies average $(\omega_-+\omega_+)/2$, which is usually close to the non-rotating disk natural angular frequency in water. The beat frequency corresponds to half the mode split magnitude given by \mbox{Eq.\ (\ref{eq:split_amp})}. Table \ref{tab:rot_water_2} compares mode split magnitudes obtained with both analytical and numerical methods for several modes; the agreeing results show how well the physics is captured. The physical interpretation of this split is that the previously stationary mode now rotates slowly at half of the mode split angular frequency magnitude $(\omega_--\omega_+)/2$. The combination of both measured frequencies $(\omega_-+\omega_+)/2$ and $(\omega_--\omega_+)/2$ allows determining the actual natural angular frequencies of the rotating disk in dense fluid:
\begin{align}
    \omega_+&=\Big(\frac{\omega_-+\omega_+}{2}\Big)-\Big(\frac{\omega_--\omega_+}{2}\Big)\,,\\
    \omega_-&=\Big(\frac{\omega_-+\omega_+}{2}\Big)+\Big(\frac{\omega_--\omega_+}{2}\Big)\,.
\end{align}
\begin{table}[ht]
    \centering
    \caption{Split magnitude of modes $n=\pm\,2,3,4$, $s=0$ obtained with the analytical model Eq.\ (\ref{eq:split_amp}) and the numerical model Eqs.\ (\ref{eq:maxZc}-\ref{eq:maxZs}) for the \citet{presas2015} rotating disk at \mbox{$\Omega_D/2\pi=40$ Hz} in water, and relative error $\epsilon$. $f=|\omega|/2\pi$.}
    \begin{tabular}{c|c|c|c}
        [Hz] & $n=2$ & $n=3$ & $n=4$\\
        \hline
        $|f_--f_+|_{,\rm{ana}}$ & 73.2 & 98.1 & 116.6\\
        $|f_--f_+|_{,\rm{num}}$ & 73.8 & 95.9 & 115.2\\
        $\epsilon_{\rm{ana-num}}$ & 0.8\% & 2.2\% & 1.2\%
    \end{tabular}
    \label{tab:rot_water_2}
\end{table}

After a certain simulated time that depends on geometrical and numerical parameters, interfering acoustic high frequency oscillations appear and prevent the observation of the beating oscillation, and therefore of the split. However, they can be avoided without modifying the disk natural frequencies by increasing the fluid compressibility, as discussed in the Appendix.

\section{Conclusion}

The analytical modal approach applied to disks gives information on the potential flow in the fluid domain above and below the plate. This further results in the determination of the \ac{AVMI} factor $\beta_0$, characterizing the fluid effect on the structural vibrations. Eventually, we determined an analytical expression for co- and counter-rotating wave angular frequencies:
\begin{equation*}
    \omega=\frac{\sqrt{(\beta_0+1)\omega_v^2-\beta_0(n\Omega_{D/F})^2}-n\beta_0\Omega_{D/F}}{\beta_0+1}\,.
\end{equation*}
We further determined that mode split and drift are respectively caused by linear and quadratic dependence of the added mass with relative wave speed. This model uses three main assumptions, namely that the disk modeshapes are unchanged by the dense fluid, that potential flow is a good approximation, and that the empirical value for the entrainment coefficient to determine the effective fluid rotation $\Omega_{D/F}$ is correct. Accepting these assumptions, the model is truly predictive, solvable in only a few seconds, and takes into account the geometry (including axial gaps, but not the radial gap), the fluid and structure characteristics, and the disk rotation. Both the frequency split and drift that result from the disk motion in dense fluid are well captured. The split amplitude asymptotic behavior is given by
\begin{equation*}
    \lim_{\beta_0\to\infty}|\omega_--\omega_+|=2n\Omega_{D/F}\,.
\end{equation*}

The modal approach applied to an arbitrary number of disk modes provides a \ac{1DOF} representation of single propagating waves. We extended this model to a 2DOF representation of companion mode pairs, namely co- and counter-rotating waves of single modes. The temporal unknowns $q_c$ and $q_s$ verify equations given by the Galerkin method resolution. We simulated freely vibrating rotating disks in dense fluid using \textsc{Ansys CFX} fluid solver coupled to the discretized equations with a Runge-Kutta method, while imposing specific modeshapes to the structure. Fluid damping and mode split are well captured. Mode split translates into a stationary mode when observed from the reference frame rotating at $(\omega_--\omega_+)/2$. Both analytical and numerical approaches agree with experimental data from \citet{presas2015}, and provide a physical interpretation of the mode split.

Although we studied the rotating disk alone, \citet{valentin2015} showed that the casing flexibility may induce coupling with the rotating disk, resulting in the modification of its natural frequencies, especially if the gaps are small. Typically, in hydraulic turbines, this issue arises when the top part of the runner vibrates with the top casing surface \citep{weder2019}. This makes rotor-stator coupling especially relevant for future work on the matter; possibly by adapting our numerical model.

Our work improves knowledge of the dynamical characteristics of high head hydraulic turbines by providing means to assess the variation of rotor natural frequencies with rotation speed and added mass of the surrounding water, which facilitates potential resonance identification within a shorter time. Our parametric and stability studies additionally show that frequency drift and flutter instability do not occur within hydraulic turbine operation range. Both models developed in this study provide fast tools for preliminary studies on high head hydraulic turbine vibrations, and ways to explore parameters influencing mode split.

\section*{Acknowledgements}

This work was supported by the MITACS Accelerate program and Andritz Hydro Canada Inc.
\newpage
\section*{Appendix: Discussion on the high frequency oscillations}

\begin{figure}[ht]
    \centering
    \includegraphics[scale=0.6]{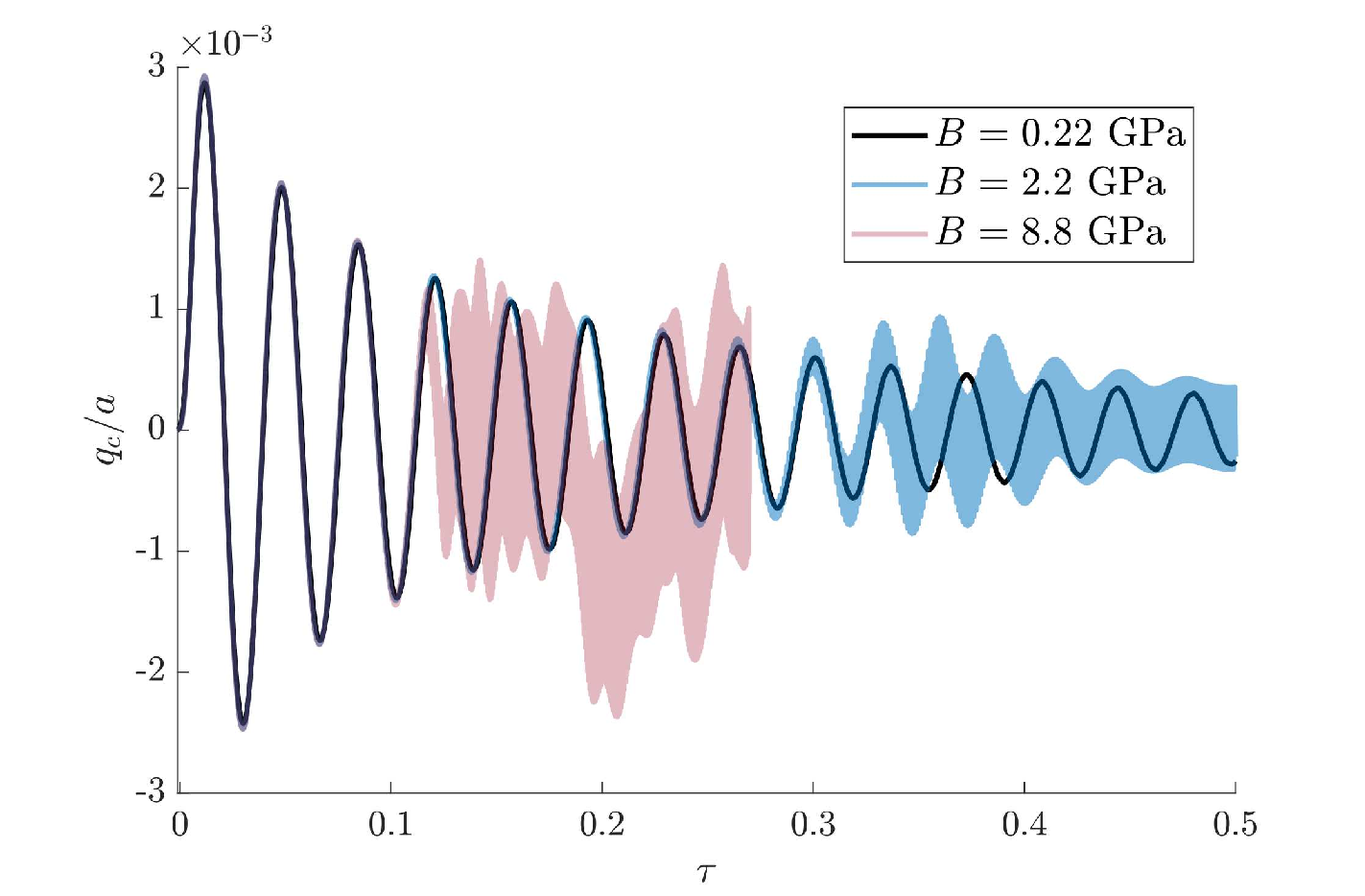}
    \caption{$q_c/a$ as a function of the elapsed dimensionless time for mode $n=3$, $s=0$ and the \citet{presas2015} geometry. The compressibility is varied through the bulk \mbox{modulus $B$}. Water corresponds to $B=2.2$ GPa. Increasing $B$ reduces the compressibility and triggers the high frequency oscillations to appear sooner. Reducing $B$ increases the compressibility and delays the high frequency oscillations. Before these appear, the $q_c/a$ signals are identical, regardless of $B$.}
    \label{fig:app_qc}
\end{figure}

\begin{figure}[ht]
    \centering
    \includegraphics[scale=0.6]{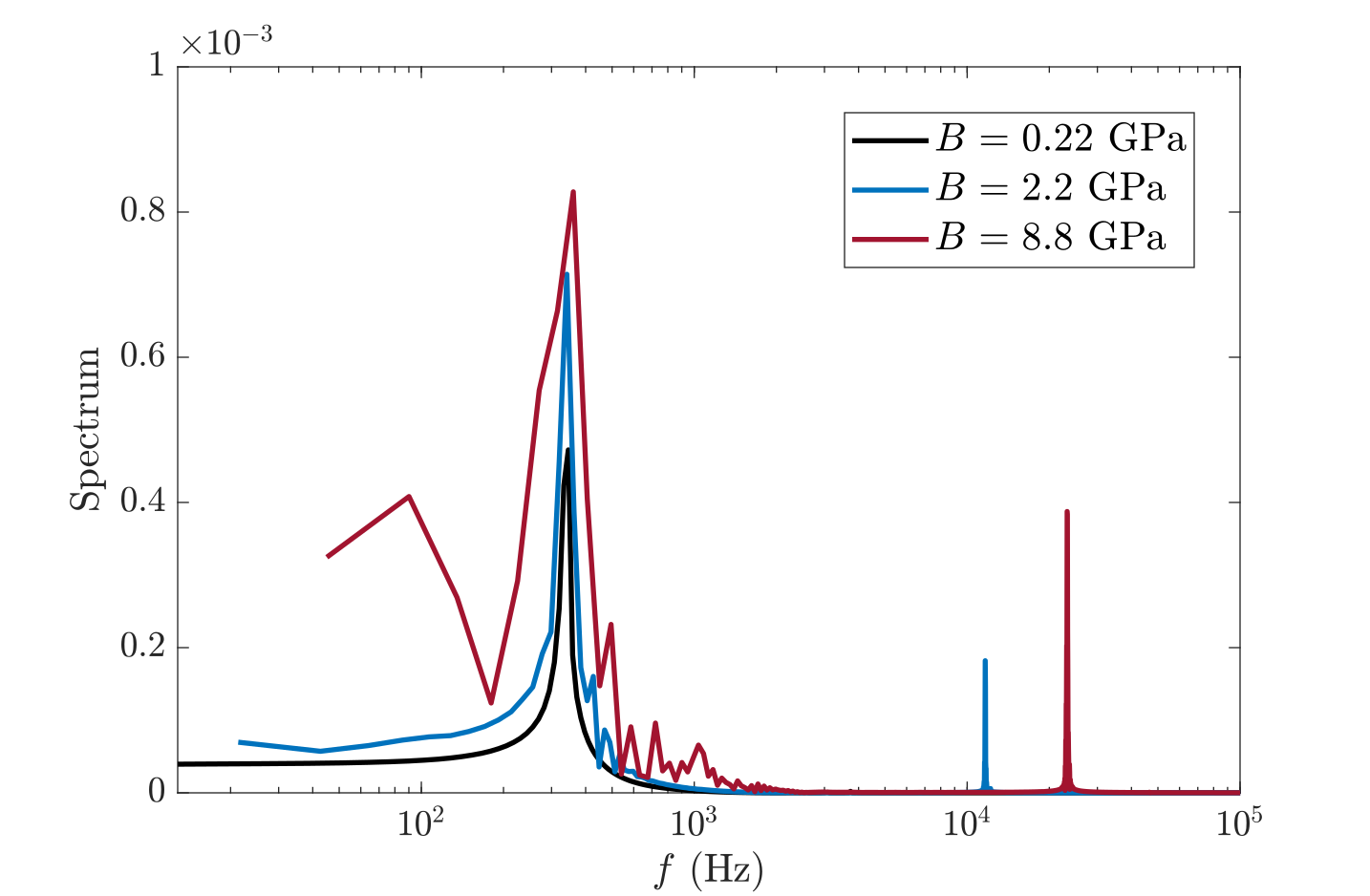}
    \caption{Fast Fourier transform of the $q_c/a$ signals shown in Figure \ref{fig:app_qc}. The low frequency corresponds to the disk structural frequency, and does not depend on the value of the bulk \mbox{modulus $B$}. The high frequency corresponds to the parasitic oscillations, and is proportional to $\sqrt{B}$.}
    \label{fig:app_fft}
\end{figure}

After a certain simulated time that depends on geometrical and numerical parameters, interfering acoustic high frequency oscillations appear and prevent the observation of the beating oscillation, and therefore of the split. Figure \ref{fig:app_qc} presents these oscillations parasitizing the $q_c/a$ signal for different fluid compressibilities. Lowering the compressibility triggers the oscillations to appear sooner in the simulation, while increasing it delays them. The signal remains identical before the oscillations appear, regardless of the compressibility. Figure \ref{fig:app_fft} shows the fast Fourier transform of these $q_c/a$ signals. The disk structural frequency does not depend on the compressibility, while the parasitic oscillations frequency is proportional to $\sqrt{B}$.

The bulk modulus $B$ is linked to the sound speed in the fluid domain $c$ by
\begin{equation*}
    c=\sqrt{B/\rho_F}\,.
\end{equation*}
This proves that the high frequency oscillations are acoustic vibrations, a different physics than for the mode split. Because the walls are perfectly reflective in the numerical model, the acoustic vibrations add up and hide the studied signal when their orders of magnitude are similar. However, because the structural frequency does not depend on $B$, choosing a low compressibility value delays the parasitic oscillations and allows the study of the $q_c/a$ and $q_s/a$ signals over sufficiently long periods of time. The black curve of Figure \ref{fig:app_qc} is a good example of how to avoid the acoustic parasitizing. Additionally using closely grouped monitor points on the disk surface also helps detect the beat in a shorter simulated physical time.

\newpage

\bibliography{mybibfile}

\end{document}